\documentclass[aps,prd,onecolumn,groupedaddress,showpacs,nofootinbib,amssymb]{revtex4}
\usepackage{graphicx}
\usepackage[english]{babel}
\usepackage{amsmath}
\usepackage{mathrsfs}
\usepackage{amssymb}
\usepackage{amsfonts}
\usepackage{epsfig}
\usepackage{colordvi}
\usepackage{color}
\usepackage{bm}

\def\de#1/de#2{\frac{\partial {#1}}{\partial {#2}}}
\def\De#1/de#2{\dfrac{\partial {#1}}{\partial {#2}}}

\def\ben{\begin{eqnarray}}
\def\een{\end{eqnarray}}
\def\be{\begin{equation}}
\def\ee{\end{equation}}
\def\ba{\begin{eqnarray}}
\def\ea{\end{eqnarray}}
\def\bea{\begin{eqnarray*}}
\def\eea{\end{eqnarray*}}

\def\umunu{^{\mu\nu}}
\def\dmunu{_{\mu\nu}}
\def\ua{^{\alpha}}
\def\ub{^{\beta}}

\def\dab{_{\alpha\beta}}

\def\beqt{\begin{tabular}}
\def\eet{\end{tabular}}
\def\beqf{\begin{figure}}
\def\eef{\end{figure}}
\def\beqa{\begin{eqnarray}}
\def\eeqa{\end{eqnarray}}

\def\gu{g^{\mu\nu}}
\def\gd{g_{\mu\nu}}

\def\umunu{^{\mu\nu}}
\def\dmunu{_{\mu\nu}}
\def\ua{^{\alpha}}
\def\ub{^{\beta}}

\def\dab{_{\alpha\beta}}

\def\pa{\partial}

\def\p{\phi}

\def\v{V(\phi)}
\def\vp{V'(\phi)}

\begin{document}
\title{Gravitationally Generated Interactions}
\author{Salvatore Capozziello$^{1,2}$, Mariafelicia De Laurentis$^{1,2}$, Luca Fabbri$^{3,4,5}$, Stefano Vignolo$^{3}$}
\affiliation{\it $^1$Dipartimento di Scienze Fisiche, Universit\`{a} di Napoli ``Federico II'', Compl. Univ. di Monte S. Angelo, Edificio G, Via Cinthia, I-80126, Napoli, Italy\\
$^2$ INFN Sez. di Napoli, Compl. Univ. di Monte S. Angelo, Edificio G, Via Cinthia, I-80126, Napoli, Italy\\
$^3$DIPTEM Sez. Metodi e Modelli Matematici, Universit\`{a} di Genova, Piazzale Kennedy, Pad. D, 16129 Genova, Italy\\
$^4$INFN Sez. di Bologna, Viale B. Pichat 6/2, I-40127 Bologna, Italy\\
$^5$ Dipartimento di Fisica, Universit\`{a} di Bologna, Via Irnerio 46, I-40126 Bologna, Italy}
\date{\today}
\begin{abstract}
Starting from a 5D-Riemannian manifold, we show that a reduction mechanism to 4D-spacetimes reproduces Extended Theories of Gravity (ETGs) that are direct generalizations of Einstein's gravity. In this context, the gravitational degrees of freedom can be dealt under the standard of spacetime deformations. Besides, such deformations can be related to the mass spectra of particles. The intrinsic non-linearity of ETGs gives an energy-dependent running coupling, while torsion gives rise to interactions among spinors displaying the structure of the weak forces among fermions. We discuss how this scheme is compatible with the known observational evidence and suggest that eventual discrepancies could be detected in experiments, as ATLAS and CMS, today running at LHC (CERN). We finally discuss the consequences of the present approach in view of unification of physical interactions.
\end{abstract}
\maketitle
\section{Introduction}
So far as we know, there are four fundamental forces in Nature: electromagnetic, weak, strong and gravitational interactions. The Standard Model represents well the first three, but not the gravitational interaction, described by the geometric theory of General Relativity (GR). 

The Standard Model of Elementary Particles can be considered a successful relativistic quantum field theory both from particle physics and group-theory point of view; technically, it is a non-Abelian gauge theory (specifically a Yang-Mills theory) associated with the tensor (Cartesian) product of the internal symmetry groups $SU(3)\times SU(2)\times U(1)$, where the $SU(3)$ color symmetry for quantum chromodynamics, is treated as exact, whereas the $SU(2)\times U(1)$ symmetry, responsible for generating the electro-weak gauge fields, is considered to be spontaneously broken. GR is built upon the Einstein equations, a non-linear set of field equations for the spacetime metric. This non-linearity is indeed a source of difficulty for the GR quantization, and with it for most of the conceptual problems that arise from having GR and quantum principles related in a unique framework. 

Since the Standard Model is a gauge theory where all the fields mediating the interactions are represented by gauge potentials, the question is why the fields mediating the gravitational interaction (i.e. the gravitational potentials) are different from those of the other fundamental forces. It is reasonable to expect that there may be a gauge theory in which the gravitational fields stand on the same footing as the other gauge fields of the Standard Model \cite{oqr}. As it is well-known, this expectation has prompted a re-examination of GR from the point of view of gauge theories: while the gauge groups involved in the Standard Model are all internal symmetry groups, the gauge groups in GR must be associated with external spacetime symmetries. Therefore, the gauge theory of gravity cannot be considered under the same standard of the Yang-Mills theories. It must be one in which gauge objects are not only gauge potentials but also tetrads that relate the symmetry group to the external spacetime. For this reason, we have to consider a more complex non-linear gauge theory where all the interactions should be dealt with under the same standards \cite{unification}. In GR, Einstein took the spacetime metric components as the basic set of variables representing gravity, whereas Ashtekar and collaborators employed the tetrad fields and the connection forms as the fundamental variables \cite{rovelli}. We also will consider the tetrads and the connection forms as the fundamental fields but with the difference that this approach gives rise to a covariant symplectic formalism capable of achieving the result of dealing with physical fields under the same standard \cite{basini,symplectism}.

In order to frame historically our approach, let us sketch a quick summary of the various attempts where the Standard Model and GR have been considered under the same comprehensive picture. In 1956, Utiyama suggested that gravitation may be viewed as a gauge theory \cite{Utiyama} in analogy to the Yang Mills \cite{YangMills} theory (1954). He identified the gauge potential due to the Lorentz group with the symmetric connection of the Riemann geometry, and reproduced the Einstein GR as a gauge theory of the Lorentz group $SO(3,1)$ with the help of tetrad fields introduced in an \textit{ad hoc} manner. Although the tetrads were necessary components of the theory (to relate the Lorentz group) adopted as an internal gauge group to the external spacetime, they were not introduced as gauge fields. In 1961, Kibble \cite{Kibble} constructed a gauge theory based on the Poincar\'{e} group $P(3,1)=T(3,1)\rtimes SO(3,1)$ (the symbol $\rtimes $ represents the semi-direct product) which resulted in the Einstein-Cartan theory characterized by curvature and torsion. The translation group $T(3,1)$ is considered responsible for generating the tetrads as gauge fields. Cartan \cite{Cartan} generalized the Riemann geometry in order to include torsion in addition to curvature. Sciama \cite{Sciama}, and others (Fikelstein \cite{Finkelstein}, Hehl \cite{Hehl1, Hehl2}) pointed out that intrinsic spin may be the source of torsion of the underlying spacetime manifold. Since the form and the role of tetrad fields are very different from those of gauge potentials, it has been thought that even Kibble's attempt was not satisfactory as a full gauge theory. There have been a number of gauge theories of gravitation based on a variety of Lie groups \cite{Hehl1, Hehl2, Mansouri1, Mansouri2, Chang, Grignani, MAG}. It was argued that a gauge theory of gravitation corresponding to GR can be constructed with the translation group alone, in the so-called teleparallel scheme \cite{teleparallelism}. Inomata \textit{et al.} \cite{Inomata} proposed that Kibble's gauge theory could be obtained, in a way closer to the Yang-Mills approach, by considering the de Sitter group $SO(4,1)$, which is reducible to the Poincar\'{e} group by a group-contraction. Unlike the Poincar\'{e} group, the de Sitter group is homogeneous and the associated gauge fields are all of gauge potential type and by the Wigner-In\"{o}nu group contraction procedure, one of the five vector potentials reduces to the tetrad. The non-linear approach to group realizations was originally introduced by Coleman, Wess and Zumino \cite{CCWZ1, CCWZ2} in the context of internal symmetry groups (1969). It was later extended to the case of spacetime symmetries by Isham, Salam, and Strathdee \cite{Isham} considering the non-linear action of $GL(4$, $\mathbf{\mathbb{R}})$, modulus the Lorentz subgroup. In 1974, Borisov, Ivanov and Ogievetsky \cite{BorisovOgievetskii, IvanovOgievetskii}, considered the simultaneous non-linear realization (NLR) of the affine and conformal groups. They stated that GR can be viewed as a consequence of spontaneous breakdown of the affine symmetry, in the same way that chiral dynamics, in quantum chromodynamics, is a result of spontaneous breakdown of chiral symmetry. In their model, gravitons are considered as Goldstone bosons associated with the affine symmetry breaking. As we will see below, this approach can be pursued in general. In 1978, Chang and Mansouri \cite{ChangMansouri} used the NLR scheme adopting $GL(4$,$\mathbf{\mathbb{R}})$ as the principal group. In 1980, Stelle and West \cite{StelleWest} investigated the NLR induced by the spontaneous breakdown of $SO(3$, $2)$. In 1982 Ivanov and Niederle considered non-linear gauge theories of the Poincar\'{e}, de Sitter, conformal and special conformal groups \cite{Ivanov1, Ivanov2}. In 1983, Ivanenko and Sardanashvily \cite{IvanenkoSardanashvily} considered gravity to be a spontaneously broken gauge theory. The tetrads fields arise, in their formulation, as a result of the reduction of the structure group of the tangent bundle from the general linear to Lorentz group. In 1987, Lord and Goswami \cite{Lord1, Lord2} developed the NLR in the fiber bundle formalism based on the bundle structure $G\left( G/H\text{,}H\right) $ as suggested by Ne'eman and Regge \cite{NeemanRegge}. In this approach, the quotient space $G/H$ is identified with physical spacetime. Most recently, in a series of papers, Lopez-Pinto, Julve, Tiemblo, Tresguerres and Mielke discussed non-linear gauge theories of gravity on the basis of the Poincar\'{e}, affine and conformal groups \cite{Julve, Lopez-Pinto, TresguerresMielke, Tresguerres, TiembloTresguerres1, TiembloTresguerres2}.

As it should be clear from this extensive resume, the idea of an unification theory, capable of describing all the fundamental interactions of physics under the same standard, has been one of the main issues of modern physics, starting from the early efforts of Einstein, Weyl, Kaluza and Klein \cite{kaluza} until the most recent approaches \cite{ross}. Nevertheless, the large number of ideas, up to now proposed, which we classify as {\it unified theories}, results unsuccessful due to several reasons: the technical difficulties connected with the lack of a unitary mathematical description of all the interactions; the huge number of parameters introduced to build the unified theory and the fact that most of them cannot be observed neither at laboratory nor at astrophysical (or cosmological) conditions \cite{morselli}; the very wide (and several times questionable since not-testable) number of extra-dimensions requested by several approaches. Due to this situation, it seems that unification is a useful (and aesthetic) paradigm, but far to be achieved, if the trend is continuing to try to unify interactions by adding more and more ingredients, rarely with an insightful justification. A different approach could be to consider the very essential physical quantities and try to achieve unification without any {\it ad hoc} new ingredients. This approach can be pursued starting from straightforward considerations which lead to reconsider modern physics under a sort of economic issue: that is, let us try to unifying forces approaching new schemes but without adding new parameters, just exploiting what already is in the underlying set-up to its full extent.

So as we said, Utiyama \cite{Utiyama} proposed that GR can be seen as a gauge theory based on the local Lorentz group in the same way that the Yang-Mills gauge theory \cite{YangMills} is developed on the basis of the internal isospin gauge group. In this formulation, the Riemannian connection is the gravitational counterpart of the Yang-Mills gauge fields. While $SU(2)$, in the Yang-Mills theory, is an internal symmetry group, the Lorentz symmetry represents the local nature of spacetime rather than internal degrees of freedom. In order to relate local Lorentz symmetry to the external spacetime, we need to solder the local space to the external space. The soldering tools can be the tetrad fields. Utiyama regarded the tetrads as objects given \textit{a priori} while they can be dynamically generated \cite{unification} and the spacetime has necessarily to be endowed with torsion in order to accommodate spinor fields \cite{classification}. In other words, the gravitational interaction of spinning particles requires the modification of the Riemann spacetime of GR to be a (non-Riemannian) curved spacetime with torsion. 
However spinor fields can be viewed as member of Clifford algebra and exist also in flat spaces \cite{kaehler}. Although Sciama used the tetrad formalism for his gauge-like handling of gravitation, his theory fell shortcomings in treating tetrad fields as gauge fields. Following the Kibble approach \cite{Kibble}, it can be demonstrated how gravitation can be formulated starting from a pure gauge viewpoint. In particular, gravity can be seen as a gauge theory which can be obtained starting from some local invariance, e.g. the local Poincar\'{e} symmetry, leading to a suitable unification prescription \cite{unification}. However, torsion is \emph{not} the minimal ingredient we will require, as it is naturally part of the most general geometry. The {\it minimal} ingredient we require is a 5-dimensional extension that, in 4-dimensions, reduces to an enlargement of the Hilbert-Einstein Lagrangian (ETGs): starting from the 5D-space, we will see that the process of reduction to 4D-space {\it generates} the mass spectra of particles, and in this picture, we will show that deformations can be parametrized as \emph{effective} scalar fields in a $GL(4)$-group of diffeomorphisms; alternatively, the Hilbert-Einstein action $R$ will be taken to be in its most general form as a generic function of the Hilbert-Einstein action $f(R)$. In this circumstance, we will see that the factor $\frac{1}{f'(R)}$ will give rise to an energy-dependent running-coupling able to bring effects that are supposed to take place at the Planck scale down to the usual Fermi scale. This is a straightforward paradigm for ETGs which could involve also other curvature invariants \cite{book,physrep}. Here, we are going to eventually see that torsion gives rise to spinorial interactions which, in some situations, can be worked out to display the same structure of the weak forces.

The layout of the paper is the following: in Section 2, we shall recall the basic geometrical concepts and clarify the notations and definitions we will employ in the rest of the paper; in Section 3, we will briefly introduce the 5D-spaces, and their reduction to 4D-spacetimes; Section 4, we discuss deformations of spacetime under the standard of $GL(4)$-group of transformations. In Section 5, we discuss the reduction from 5D to 4D showing how massive states can come out. Section 6 is devoted to the generation of masses and their relation to geometrical deformations. In Section 7, we are going to deal with the concept of mass generation and symmetry breaking. The further gravitational degrees of freedom, coming from ETGs, can play a fundamental role in this picture generating natural cut-off at TeV scales: this fact could give rise to gravitationally induced electroweak interactions. In particular, as discussed in Section 8, the weak forces emerge from the interaction of torsion and spinors and a running coupling constant is directly related to the form of $f(R)$-gravity. A detailed discussion of this feature is reported. Conclusions are drawn in Section 9.
\section{Geometrical Notations and Definitions}
To begin with, we shall recall some concepts and introduce the notation we will follow. 
The fundamental objects we are going to work with are tensors, defined in terms of their transformation laws, and whose indices, upper or lower, may be converted into one another, by raising lower or lowering upper indices, by means of the tensors $g^{\alpha\beta}$ and $g_{\alpha\beta}$ respectively; in order to ensure that the raising and lowering of the indices is consistently defined, these two tensors are assumed to be non-degenerate, symmetric and one the inverse of the other $g_{\nu\rho}g^{\rho\mu}=\delta_{\nu}^{\mu}$, therefore displaying all properties that make them metric tensors.

Dynamical properties of tensors are defined through covariant derivatives $D_{\mu}$ which are themselves defined upon introduction of the connections $\Gamma^{\alpha}_{\mu\nu}$ defined in term of its transformation law alone, and because the most general connection is not symmetric in the two lower indices then its antisymmetric part does not vanish and writing it as
\begin{equation}
T^{\lambda}_{\phantom{\lambda}\mu\nu}=\Gamma^{\lambda}_{\phantom{\lambda}[\mu,\nu]}
\end{equation}
it turns out to be a tensor called Cartan torsion tensor; in order to ensure that the raising and lowering of the indices is consistently defined even after differential properties are introduced, these two tensors are assumed vanish their covariant derivatives $D_{\mu}g=0$, and therefore we will say that the metric tensors are covariantly constant or verifying metricity or again that the metric and connection are compatible: after this condition is taken into account, the most general connection can be decomposed as
\begin{eqnarray}
&\Gamma^{\mu}_{\phantom{\mu}\sigma\pi}
=\frac{1}{2}g^{\mu\rho}\left(\partial_{\pi}g_{\sigma\rho}
+\partial_{\sigma}g_{\pi\rho}-\partial_{\rho}g_{\sigma\pi}\right)
+\left(T^{\mu}_{\phantom{\mu}\sigma\pi}
+T_{\sigma\pi}^{\phantom{\sigma\pi}\mu}+T_{\pi\sigma}^{\phantom{\pi\sigma}\mu}\right)
\end{eqnarray}
where
\begin{eqnarray}
&K^{\mu}_{\phantom{\mu}\sigma\pi}=T^{\mu}_{\phantom{\mu}\sigma\pi}
+T_{\sigma\pi}^{\phantom{\sigma\pi}\mu}+T_{\pi\sigma}^{\phantom{\pi\sigma}\mu}
\end{eqnarray}
is defined as the contorsion tensor, and which may turn out to be useful in some context.

Due to their symmetry properties, torsion and contorsion have only one independent component $K^{\mu}=K^{\mu\pi}_{\phantom{\mu\pi}\pi}=2T_{\pi}^{\phantom{\pi}\pi\mu}=2T^{\mu}$ and called Cartan torsion or contorsion vectors.

The quantity be expressed as
\begin{equation}
R^{\rho}_{\phantom{\rho}\sigma\mu\nu}=\partial_{\nu}\Gamma^{\rho}_{\phantom{\rho}\sigma\mu}
-\partial_{\mu}\Gamma^{\rho}_{\phantom{\rho}\sigma\nu}
+\Gamma^{\rho}_{\phantom{\rho}\lambda\nu}\Gamma^{\lambda}_{\phantom{\lambda}\sigma\mu}
-\Gamma^{\rho}_{\phantom{\rho}\lambda\mu}\Gamma^{\lambda}_{\phantom{\lambda}\sigma\nu}
\label{curvature}
\end{equation}
is a tensor antisymmetric in the first and second couple of indices called Riemann curvature tensor; notice that the first couple of indices is antisymmetric only after the metricity condition is accounted: this expression of the Riemann curvature tensor has been given so that, together with the expression of the Cartan torsion tensor, we may now write the commutator of two covariant derivatives as
\begin{eqnarray}
&[D_{\mu},D_{\nu}]A^{\rho}=T^{\theta}_{\phantom{\lambda}\mu\nu}D_{\theta}A^{\rho}
-\frac{1}{2}R^{\rho}_{\phantom{\rho}\sigma\mu\nu}A^{\sigma}
\end{eqnarray}
for the case of a generic vector $A^{\mu}$, and for which it is understood that for general tensor there will appear an additional contribution of the Riemann curvature tensor for any additional index. In particular for scalars, the Riemann curvature tensor would not be present at all, thus underlining the specific role of the Cartan torsion tensor in describing the failure of the commutation of the covariant derivatives even for scalar fields.

Due to their symmetry properties, curvature has only one independent component that is given by $R^{\rho}_{\phantom{\rho}\sigma\rho\nu}=R_{\sigma\nu}$ called Ricci curvature tensor, with one component $R_{\sigma\nu}g^{\sigma\nu}=R$ called Ricci curvature scalar.

The commutator of three coordinate covariant derivatives in cyclic permutations gives the expressions
\begin{eqnarray}
\nonumber
&(2D_{\kappa}T^{\rho}_{\phantom{\rho}\mu \nu}
+4T^{\rho}_{\phantom{\rho}\kappa \pi}T^{\pi}_{\phantom{\pi}\mu \nu}
-R^{\rho}_{\phantom{\rho}\kappa\mu\nu})
+(2D_{\nu}T^{\rho}_{\phantom{\rho} \kappa \mu}
+4T^{\rho}_{\phantom{\rho}\nu \pi}T^{\pi}_{\phantom{\pi}\kappa \mu}
-R^{\rho}_{\phantom{\rho}\nu\kappa\mu})+\\
&+(2D_{\mu}T^{\rho}_{\phantom{\rho} \nu \kappa}
+4T^{\rho}_{\phantom{\rho}\mu \pi}T^{\pi}_{\phantom{\pi} \nu \kappa}
-R^{\rho}_{\phantom{\rho}\mu\nu\kappa})\equiv0
\end{eqnarray}
known as the Bianchi identities for torsion and
\begin{eqnarray}
\nonumber
&(D_{\mu}R^{\nu}_{\phantom{\nu}\iota \kappa \rho}
-2R^{\nu}_{\phantom{\nu}\iota \beta \mu}T^{\beta}_{\phantom{\beta}\kappa\rho})
+(D_{\kappa}R^{\nu}_{\phantom{\nu}\iota \rho \mu}
-2R^{\nu}_{\phantom{\nu}\iota \beta \kappa}T^{\beta}_{\phantom{\beta}\rho\mu})+\\
&+(D_{\rho}R^{\nu}_{\phantom{\nu}\iota \mu \kappa}
-2R^{\nu}_{\phantom{\nu}\iota \beta \rho}T^{\beta}_{\phantom{\beta}\mu\kappa})\equiv0
\end{eqnarray}
known as Bianchi identities for curvature. 

Of course, these identities can be contracted, and in their fully contracted form they read as in the following
\begin{eqnarray}
&D_{\rho}\left(T^{\rho}_{\phantom{\rho}\mu \nu} +\delta^{\rho}_{\nu}T_{\mu}-\delta^{\rho}_{\mu}T_{\nu}\right)
+2T_{\rho}\left(T^{\rho}_{\phantom{\rho}\mu \nu} +\delta^{\rho}_{\nu}T_{\mu}-\delta^{\rho}_{\mu}T_{\nu}\right)+R_{[\mu\nu]}\equiv0
\end{eqnarray}
known as fully contracted Bianchi identities for torsion and
\begin{eqnarray}
&D_{\nu}\left(R^{\nu\rho}-\frac{1}{2}g^{\nu\rho}R\right)
-2R_{\nu\beta}T^{\beta\nu\rho}+T_{\beta\mu\nu}R^{\mu\nu\beta\rho}\equiv0
\end{eqnarray}
known as fully contracted Bianchi identities for curvature.

A final remark is to be made for torsion: because torsion is a tensor, then it may be separated away, so that all most general torsional quantities can be written in terms of simplest torsionless quantities plus torsional corrections. In the following, we will employ the symbols $\nabla$ and $\mathcal{R}$ to designate the covariant derivatives and curvatures of the simplest torsionless connection.

Now, given the metric tensors $g_{\alpha\beta}$ and $g^{\alpha\beta}$ it is possible to define covariant metric concepts like lengths and angles; let us considered a pair of bases of vectors called beins $V^{a}_{\beta}$ and $V_{a}^{\beta}$ defined to be dual of one another $V^{a}_{\mu}V_{a}^{\rho}=\delta^{\rho}_{\mu}$ and $V^{a}_{\mu}V_{r}^{\mu}=\delta_{r}^{a}$ which we can always choose to be orthonormal $V^{a}_{\sigma}V^{b}_{\rho}g^{\sigma\rho}=\eta^{ab}$ or equivalently $V_{a}^{\sigma}V_{b}^{\rho}g_{\sigma\rho}=\eta_{ab}$ where $\eta_{ab}$ and $\eta^{ab}$ are unitary diagonal matrices known as Minkowskian matrices: although it is always possible to orthonormalize a basis of vectors so that it is without loss of generality that orthonormal beins are introduced, nevertheless the bein are determined up to a Lorentz transformation that can be made explicit. The main advantage of introducing the beins is that, as the metric tensors are used to move indices up or down in tensors, the bein are used to switch to spacetime Greek indices to Lorentz Latin indices, whose explicit transformation laws allow for a simpler description of the formalism; for the moment, this formalism will only give us a simpler way to deal with calculations, but in the following we will see it will be necessary for the introduction of the spinor fields.

According to the definition of bein then, the world tensors are defined in terms of their transformation law under Lorentz transformations, whose indices, upper or lower, may be converted into one another, by raising lower or lowering upper indices, by means of the Minkowskian matrices, symmetric and inverse of one another $\eta_{ab}\eta^{bm}=\delta_{a}^{m}$ as expected. 

Dynamical properties of tensors are defined by covariant derivatives $D_{\mu}$ which are themselves defined upon introduction of the spin-connections $A^{i}_{j\mu}$ which is a connection under Lorentz transformation although it is a covector under general coordinate transformations; if we want that the passage between spacetime and Lorentz indices is consistent even after differential properties are introduced, then we have to assume the condition $D_{\alpha}V=0$ analogously to what we have done before, and therefore the intuitive condition $D_{\alpha}\eta=0$ is given automatically: these two conditions together imply that the spin-connection is given by
\begin{eqnarray}
&A^{b}_{\phantom{b}j\mu}=
V^{\alpha}_{j}V_{\rho}^{b}\left(\Gamma^{\rho}_{\phantom{\rho}\alpha\mu}
+V_{\alpha}^{k}\partial_{\mu}V^{\rho}_{k}\right)
\label{Lorentzconnection}
\end{eqnarray}
in terms of the connection and antisymmetric in the two world indices.

Considering the spin-connection it is possible to define
\begin{eqnarray}
&R^{a}_{\phantom{a}b\sigma\pi}
=\partial_{\pi}A^{a}_{\phantom{a}b\sigma}-\partial_{\sigma}A^{a}_{\phantom{a}b\pi}
+A^{a}_{\phantom{a}j\pi}A^{j}_{\phantom{j}b\sigma}
-A^{a}_{\phantom{a}j\sigma}A^{j}_{\phantom{j}b\pi}
\end{eqnarray}
antisymmetric in both spacetime and world indices and such that we have the simple relationship $R^{a}_{\phantom{a}b\sigma\pi}=R^{\mu}_{\phantom{\mu}\rho\sigma\pi}V^{\rho}_{b}V^{a}_{\mu}$ in terms of the Riemann curvature tensor, as it should have been obvious since all definitions have been given so to reduce the passage from the two formalisms a mere renaming of indices; in this formalism, the antisymmetry in the two Lorentz indices is manifest. The commutators are given in an equivalent way.

As we have anticipated, the introduction of the bein are able to simplify the formalism, but their most important contribution is the fact that through them we are capable of defining new fields such as the spinors. To see that, it is enough to remark that, whereas nothing can be done for general coordinate transformations, the Lorentz transformations have an explicit structure, which can be written in different representations; of all these representations, the most special are the complex-valued representations: complex representations of the Lorentz transformations are called spinorial transformations $S$ and they can be expanded in terms of their infinitesimal generators
\begin{eqnarray}
&\gamma_{ij}=\frac{1}{4}[\gamma_{i},\gamma_{j}]
\end{eqnarray}
where the $\gamma_{a}$ matrices belong to the Clifford algebra.

Once the spinorial transformation $S$ is obtained, we define the complex fields that transform according to this transformation as spinor fields, classified in terms of half-integer spin; for what we will have to do, we will restrict ourselves to the case given by the fields transforming as $\psi'=S\psi$ or $\overline{\psi}'=\overline{\psi}S^{-1}$, being $\frac{1}{2}$-spin spinor fields, for which the passage between the two forms is defined by the $\gamma_{0}$ matrix as $\overline{\psi}\equiv\psi^{\dagger}\gamma_{0}$ or $\psi\equiv\gamma_{0} \overline{\psi}^{\dagger}$ reciprocally.

Dynamical properties for spinor fields are defined through spinorial covariant derivatives, after the introduction of the spinorial connection $\Omega_{\mu}$ as usual; we notice that for the way in which they have been built, the $\gamma_{j}$ and $\gamma_{ij}$ matrices are constant for the spinorial covariant derivatives automatically: consequently, we have that the most general spinorial connection can be decomposed as
\begin{eqnarray}
&\Omega_{\mu}=A^{ab}_{\phantom{ab}\mu}\gamma_{ab}+iqA_{\mu}\mathbb{I}
\end{eqnarray}
in terms of the spin-connection plus an additional term proportional to the identity matrix that will be identified with the electrodynamic potential, as we shall see in a moment.

Given the spinorial connection it is possible to define
\begin{eqnarray}
&R_{\sigma\pi}=\partial_{\sigma}\Omega_{\pi}-\partial_{\pi}\Omega_{\sigma}
+\Omega_{\sigma}\Omega_{\pi}-\Omega_{\pi}\Omega_{\sigma}
\end{eqnarray}
antisymmetric and such that $R_{\sigma\pi}=R^{ij}_{\phantom{ij}\sigma\pi}\gamma_{ij}+iqF_{\sigma\pi}\mathbb{I}$ in terms of the Riemann and Maxwell tensors. Eventually, the commutators will equivalently be defined as above.

It is important to notice that for the decomposition of the most general spinorial connection, the additional term proportional to the identity matrix should not be surprising: in fact, because spinors are complex fields, then we have to consider an additional transformation for a complex unitary phase $\psi'=e^{iq\theta}\psi$ or $\overline{\psi}'=e^{-iq\theta}\overline{\psi}$ for a given real number $q$ called charge, and therefore the issue of their invariance under gauge transformations is to be considered as well; the theory that is thus constructed is the widely known electrodynamic Maxwell theory. This abelian gauge potential will eventually be placed in the spinorial connection; and the fact that the spinorial connection already had enough room to host this abelian gauge field should therefore not be so surprising any longer. In this sense then, electrodynamics has the same mathematical imprint of gravitation, and likewise it should be thought as a genuine geometrical theory.

The reason for which usually it is not can be traced back to a little difference we already underlined, that is the presence of the charge; however, the charge labels a property of the matter field, and therefore what is not geometric is the individual response of a given matter field propagating in an assigned gauge field, but the electrodynamic field in itself might indeed be considered as a purely geometrical field.

This concludes the introduction of the formalism, through which we not only defined the fundamental quantities we will employ in the following, but we also defined them in order to underline the geometrical essence beneath: this does not only refer to the well-known geometrical essence of the gravitational field, but also to the geometrical essence of the electrodynamic field (at least in terms of the field itself and not the response of a matter field in it), and the geometric settings of the spinorial fields (before field equations are defined); these ideas are actually not new, and for all concepts recalled here, a general introduction as well as a deeper discussion about the idea of geometrization, can be found in \cite{s-s} (additionally it is possible to find references about all different roles of torsion).

We notice that so far, the formalism we have introduced is still non-dynamical (we have not defined any field equation or action) and still valid in any dimension (we have not chosen a particular background); in the following, we will discuss generalized actions for gravity. Specifically, we will take into account a 5D-spacetime with dynamical reduction to 4D-spacetime.
\section{The geometrical structure of 5D and 4D spaces}
In this section, we will discuss the structure of a 5D-Riemannian manifold, and its reduction to 4D-manifold: such an approach gives a useful tool to deal with the realization of effective theories of gravity in 4D and the problem of mass generation. 

Let us start with the 5D-space, where we do not define {\it a priori} a signature for the metric and which, after a 4D-reduction procedure, must be capable of reproducing all the features of the usual spacetime \cite{sabafe}. Let us assume that a 5D-vector field defines a metric whose
signature is given by 
\begin{eqnarray}
&(X,X)=(x^0)^2-\overrightarrow{X}\cdot\overrightarrow{X}+\epsilon (x_4)^2\,
\end{eqnarray}
where $\epsilon=\pm 1$, so that, using the traditional terms, the fifth dimension can be time-like or space-like: as we shall see below, it is the 4D-dynamics that discriminates, by a bijective correspondence, the signature, giving rise to particle-like solutions for $\epsilon=-1$ or to
wave-like solutions for $\epsilon=+1$. In this case, we can obtain pseudo-spheres of Lorentz signature and thus 4D-spacetimes of constant curvature (as Friedmann-Robertson-Walker ones). There
are only two independent different signatures for $N=5$: they are $(p,q)=(1,4)$, corresponding to the case $\epsilon=-1$ and $(p,q)=(2,3)$, corresponding to $\epsilon=+1$. The 5D-manifolds are ${\cal M}^{(1,4)}={\bf R}^5$ and ${\cal M}^{(2,3)}={\bf R}^5$ respectively, where ${\bf R}^5$ is the 5D-space, the former case is called the De Sitter space, while the latter is the Anti-De Sitter one. The fact that the standard signature of the universe is $(1,-1,-1,-1)$ can be derived from an equivalent process starting from ${\cal M}^{(1,4)}$ or ${\cal M}^{(2,3)}$. This procedure is in agreement with the Whitney Theorem \cite{skopenkov}.

An important consideration is necessary at this point. After the reduction to 4D-physics, a given theory of gravity is described by the conformal-affine group of diffeomorphisms $GL(4)$ which is characterized by $4\times 4=16$ generators; a straightforward splitting is
\begin{equation}
\underbrace{GL(4)}_{\underbrace{4\times 4}}\supset \underbrace{SU(3)}_{\underbrace{3^2-1}} \otimes \underbrace{SU(2)}_{\underbrace{2^2-1}} \otimes \underbrace{U(1)}_{\underbrace{1}}\otimes\underbrace{ GL(2)}_{\underbrace{2\times 2}}
\end{equation}
where the number of generators is indicated for any subgroup: the physical meaning of $SU(3)$, $SU(2)$ and $U(1)$ is clear while $GL(2)$ is a group with 4 generators.

The further generators can be recovered in the framework of ETGs being massive or ghost gravitational modes \cite{book,physrep,odino,sabafe}. Clearly the groups of diffeomorphisms $GL(4)$ and $GL(2)$ are not unitary and this could be considered a problem in the standard approaches to quantization. On the contrary, from our point of view, this feature are related to mechanisms suitable to induce electroweak interactions and gravitational running coupling constants. The experimental consequences of these considerations could be extremely interesting for the physics at LHC.
\section{Deformations and the Physics of the GL(4)-group}
Another ingredient for our considerations is represented by the spacetime deformations that can be induced in the reduction mechanism and can be seen as scalar fields in a gravitational background \cite{deform}. Let us take into account a metric $\mathbf{g}$ on a spacetime manifold $\mathcal{M}$ (in $N$-dimensions); let us now define a new tetrad field $\widetilde{\omega}=\Phi^{A}_{\phantom{A}C}(x)\,\omega^{C}$, with $\Phi^{A}_{\phantom{A}C}(x)$ a matrix of scalar fields, and finally introduce a spacetime $\widetilde{\mathcal{M}}$ with the metric $\widetilde{g}$ defined as
\begin{equation}
\label{deformed}
\mathbf{\widetilde{g}}=\eta_{AB}\Phi^{A}_{\phantom{A}C}\Phi^{B}_{\phantom{B}D}\,\omega^{C}
\omega^{D} = \gamma_{CD}(x)\omega^{C}
\omega^{D},
\end{equation}
where also $\gamma_{CD}(x)$ is a matrix of fields which are scalars with respect to the coordinate transformations.

If $\Phi^{A}_{\phantom{A}C}(x)$ is a Lorentz matrix in any point of $\mathcal{M}$, then $\widetilde{g}\equiv g$, otherwise we say that $\widetilde{g}$ is a deformation of $g$ and $\widetilde{\mathcal{M}}$ is a deformed $\mathcal{M}$. If all the functions of $\Phi^{A}_{\phantom{A}C}(x)$ are continuous, then there is a one-to-one correspondence between the points of $\mathcal{M}$ and the points of $\widetilde{\mathcal{M}}$. If $\xi$ is a Killing vector for $g$ on $\mathcal{M}$, the corresponding vector $\widetilde{\xi}$ on $\widetilde{\mathcal{M}}$ is not a Killing vector.

A particular subset of these deformation matrices is given by
\begin{equation}
\label{conformal}
{\Phi}{^{A}_{C}}(x)=\Omega(x)\, \delta^{A}_{\phantom{A}C}.
\end{equation}
which define conformal transformations of the metric $\widetilde{g}=\Omega^{2}(x) g$. In this sense, the deformations defined by Eq. (\ref{deformed}) can be regarded as a generalization of the conformal transformations.

We call the matrices $\Phi^{A}_{\phantom{A}C}(x)$ First Deformation Matrices, while we can refer to
\begin{equation}
\label{seconddeformation}
\gamma_{CD}(x)=\eta_{AB}\Phi^{A}_{\phantom{A}C}(x)\Phi^{B}_{\phantom{B}D}(x).
\end{equation}
as the Second Deformation Matrices which, as seen above, are also matrices of scalar fields. They generalize the Minkowski matrix \(\eta_{AB}\) with constant elements in the definition of the metric. A further restriction on the matrices $\Phi^{A}_{\phantom{A}C}$ comes from the above mentioned theorem proved by Riemann for which an $N$-dimensional metric has $N(N-1)/2$ degrees of
freedom (see \cite{Coll:2001wy} for details). With this definitions in mind, let us consider the main properties of deforming matrices. Let us take into account a four dimensional spacetime with
Lorentzian signature. A family of matrices $\Phi^{A}_{\phantom{A}C}(x)$ such that
\begin{equation}
\label{fi}
\Phi^{A}_{\phantom{A}C}(x)\in GL(4)\, \forall x,
\end{equation}
are defined on such a spacetime. These functions are not necessarily continuous and can connect
spacetimes with different topologies. A singular scalar field introduces a deformed manifold $\widetilde{\mathcal{M}}$ with a spacetime singularity.

As it is well known, the Lorentz matrices $\Lambda^{A}_{\phantom{A}C}$ leave the Minkowski metric invariant and then
\begin{equation}
\label{deformlambda2}
\mathbf{g}=\eta_{EF}\Lambda^{E}_{\phantom{E}A}\Lambda^{F}_{\phantom{F}B}\Phi^{A}_{\phantom{A}C}\Phi^{B}_{\phantom{B}D}\,
\omega^{C}
\omega^{D} =\eta_{AB}\Phi^{A}_{\phantom{A}C}\Phi^{B}_{\phantom{B}D}\,\omega^{C}
\omega^{D}
\end{equation}
and it follows that $\Phi^{A}_{\phantom{A}C}$ give rise to right cosets of the Lorentz group, {\it i.e.} they are the elements of the quotient group $GL(4,\mathbf{R})/SO(3,1)$. On the other hand, a
right-multiplication of $\Phi^{A}_{\phantom{A}C}$ by a Lorentz matrix induces a different deformation matrix.

The inverse deformed metric is
\begin{equation}
\label{inversemetric}
\widetilde{g}^{\alpha\beta}=\eta^{CD}{\Phi^{-1}}^{A}_{\phantom{A}C}
{\Phi^{-1}}^{B}_{\phantom{B}D}e_{A}^{\alpha}e_{B}^{\beta}
\end{equation}
where ${\Phi^{-1}}^{A}_{\phantom{A}C}{\Phi}^{C}_{\phantom{C}B}=\Phi^{A}_{\phantom{A}C}{\Phi^{-1}}^{C}_{\phantom{C}B}=\delta^{A}_{B}$ and we can decompose the matrix $\Phi_{AB}=\eta_{AC}\, \Phi^{C}_{\phantom{C}B}$ in its symmetric and antisymmetric parts
\begin{equation}
\label{decomposition1}
\Phi_{AB}= \Phi_{(AB)}+\Phi_{[AB]}= \Omega\,\eta_{AB} + \Theta_{AB} + \phi_{AB}
\end{equation}
where $ \Omega= \Phi^{A}_{\phantom{A}A}$, $\Theta_{AB}$ is the traceless symmetric part and $\phi_{AB}$ is the antisymmetric part of the first deformation matrix. Then standard conformal transformations are nothing else but deformations with $\Theta_{AB}=\phi_{AB}=0$ \cite{wald}.

Finding the inverse matrix ${\Phi^{-1}}^{A}_{\phantom{A}C}$ in terms of $\Omega$, $\Theta_{AB}$ and $\phi_{AB}$ is not immediate, but as above, it can be split in the three terms
\begin{equation}
\label{inversesplitting}
{\Phi^{-1}}^{A}_{\phantom{A}C}=\alpha\delta^{A}_{\phantom{A}C}+\Psi^{A}_{\phantom{A}C}+\Sigma^{A}_{\phantom{A}C}
\end{equation}
where $\alpha$, $\Psi^{A}_{\phantom{A}C}$ and $\Sigma^{A}_{\phantom{A}C}$ are respectively the trace, the traceless symmetric part and the antisymmetric part of the inverse deformation matrix above. The second deformation matrix, from the above decomposition, takes the form
\begin{equation}
\label{secondmatrix}
\gamma_{AB}= \eta_{CD}(\Omega\, \delta_{A}^{C}+
\Theta_{\phantom{C}A}^{C}+ \phi_{\phantom{C}A}^{C})(\Omega\, \delta_{B}^{D}+
\Theta_{\phantom{D}B}^{D}+ \phi_{\phantom{D}B}^{D})
\end{equation}
and then
\begin{eqnarray}
\label{secondmatrix1}
\gamma_{AB}&=& \Omega^{2}\,\eta_{AB} + 2\Omega\,\Theta_{AB}+ \eta_{CD}\, \Theta_{\phantom{C}A}^{C}\,\Theta_{\phantom{D}B}^{D} +\nonumber\\&&+ \eta_{CD}\, (\Theta_{\phantom{C}A}^{C}\,\phi_{\phantom{D}B}^{D}
+\phi_{\phantom{C}A}^{C}\,\Theta_{\phantom{D}B}^{D}) + \eta_{CD}\,\phi_{\phantom{C}A}^{C}\,\phi_{\phantom{D}B}^{D}
\end{eqnarray}
as it is clear.

In general, the deformed metric can be split as
\begin{equation}
\label{splits}
{\tilde{g}}{_{\alpha\beta}}=\Omega^{2}{g}{_{\alpha\beta}}+{\gamma}{_{\alpha\beta}}
\end{equation}
where
\begin{eqnarray}
\label{acca}
{\gamma}{_{\alpha\beta}}&=&\left( 2\Omega\,\Theta_{AB}+ \eta_{CD}\, \Theta_{\phantom{C}A}^{C}\,\Theta_{\phantom{D}B}^{D} + \eta_{CD}\, (\Theta_{\phantom{C}A}^{C}\,\phi_{\phantom{D}B}^{D}+\right. \nonumber\\ && \left. + \phi_{\phantom{C}A}^{C}\,\Theta_{\phantom{D}B}^{D})+ \eta_{CD}\,\phi_{\phantom{C}A}^{C}\,\phi_{\phantom{D}B}^{D}\right){\omega}{^{A}_{\alpha}}{\omega}{^{B}_{\beta}}
\end{eqnarray}
and in particular, if $\Theta_{AB}=0$, the deformed metric simplifies to
\begin{equation}
\label{split}
\widetilde{g}_{\alpha\beta}=\Omega^{2}g_{\alpha\beta}+\eta_{CD}\,\phi_{\phantom{C}A}^{\,C}\,\phi_{\phantom{D}B}^{\,D}
\omega^{A}_{\phantom{A}\alpha}\omega^{B}_{\phantom{B}\beta}
\end{equation}
so that, for $\Omega=1$, the deformation of a metric consists in adding to the background metric a tensor $\gamma_{\alpha\beta}$. We have to remember that all these quantities are not independent as, by the theorem mentioned in \cite{Coll:2001wy}, they have to form at most six independent functions in a four dimensional spacetime. Similarly the controvariant deformed metric can be always decomposed in the following way
\begin{equation}
\label{controvariantdecomposition}
\widetilde{g}^{\alpha\beta}= \alpha^{2}g^{\alpha\beta}+ \lambda^{\alpha\beta}
\end{equation}
identically. To find the relation between ${\gamma}{_{\alpha\beta}}$ and $\lambda^{\alpha\beta}$ we may use $\widetilde{g_{\alpha\beta}}\widetilde{g^{\beta\gamma}}=\delta_{\alpha}^{\gamma}$, and then
\begin{equation}\label{relationgammalambda}
\alpha^{2}\Omega^{2}\delta_{\alpha}^{\gamma}+ \alpha^{2}\gamma_{\alpha}^{\gamma}+\Omega^{2}\lambda_{\alpha}^{\gamma}+
{\gamma}{_{\alpha\beta}}\lambda^{\beta\gamma}=\delta_{\alpha}^{\gamma}
\end{equation}
and if the deformations are conformal transformations, we have $\alpha=\Omega^{-1}$, so that assuming such a condition, one obtains the following matrix equation
\begin{equation}
\label{relationgammalambda1}
\alpha^{2}\gamma_{\alpha}^{\gamma}+\Omega^{2}\lambda_{\alpha}^{\gamma}+
{\gamma}{_{\alpha\beta}}\lambda^{\beta\gamma}=0\,,
\end{equation}
and
\begin{equation}\label{lambdaaa}
(\delta_{\alpha}^{\beta}+
\Omega^{-2}{\gamma}_{\alpha}^{\beta})\lambda_{\beta}^{\gamma}=-\Omega^{-4}\gamma_{\alpha}^{\gamma}
\end{equation}
and finally
\begin{equation}\label{lambdaaaa}
\lambda_{\beta}^{\gamma}=-\Omega^{-4}{{(\delta+
\Omega^{-2}{\gamma})^{-1}}}{^{\alpha}_{\beta}}\gamma_{\alpha}^{\gamma}
\end{equation}
where ${(\delta+ \Omega^{-2}{\gamma})^{-1}}$ is the inverse tensor of $(\delta_{\alpha}^{\beta}+ \Omega^{-2}{\gamma}_{\alpha}^{\beta})$. 

To each matrix $\Phi^{A}_{\phantom{A}B}$ we can associate a $(1,1)$-tensor $\phi^{\alpha}_{\phantom{\alpha}\beta}$ defined by
\begin{equation}\label{3.1}
\phi^{\alpha}_{\phantom{\alpha}\beta}= \Phi^{A}_{\phantom{A}B}\omega^{B}_{\beta}e_{A}^{\alpha}
\end{equation}
such that
\begin{equation}\label{3.2}
\widetilde{g}_{\alpha\beta}=g_{\gamma\delta}\phi^{\gamma}_{\phantom{\gamma}\alpha}\phi^{\delta}_{\phantom{\delta}\beta}
\end{equation}
and viceversa from a $(1,1)$-tensor $\phi^{\alpha}_{\phantom{\alpha}\beta}$, we can define a matrix of scalar fields as
\begin{equation}\label{3.3}
\phi^{A}_{\phantom{A}B}=\phi^{\alpha}_{\phantom{\alpha}\beta} \omega_{\alpha}^{A}e_{B}^{\beta}
\end{equation}
as it is to be expected.

In conclusion, the spacetime deformations are endowed with the conformal-affine structure of $GL(N)$ groups (in particular $GL(4)$) and passing from a given metric to another gives rise to induced scalar fields that, as we shall show below, can be generated by a dimensional reduction mechanism.
\section{The 5D-space and its reduction to 4D-spacetime dynamics}
We are interested in a reduction mechanism capable of generating the masses of particles and a symmetry breaking capable of generating the observed interactions. To achieve these goals, we start from a theory of gravity, formulated in 5D and analyze its reduction in 4D. The results of this approach are ETGs. In the following, 5D indices will be capitalized. We can start from the matter-free case in 5D-gravity. As it turns out, the absence of the Dirac field gives rise to a torsion-free solution: the structure of the field equations thus reduce to that of the usual Einstein gravity. However, it is possible to exploit the fact that the 5D-space gives rise, after the reduction to the 4D-spacetime, to additional degrees of freedom that can be physically interpreted: these fields arise from deformations as new scalar fields. 

Now, in 5D-spaces, the field equations are derived from the 5D Hilbert-Einstein action as discussed above; its variation gives the $5$-dimensional field equations in the vacuum
\begin{equation}
\label{a9} 
G_{AB}=\mathcal{R}_{AB}-\frac{1}{2}g_{AB}\mathcal{R}=0
\end{equation}
where we have defined for simplicity the tensor $G_{AB}$ known as Einstein tensor: here the Ricci-flat space is always a solution. 

The 5D energy tensor for scalar fields is defined as usual
\begin{equation}
\label{b11} T_{AB}=\nabla_{A}\Phi\nabla_{B}\Phi-\frac{1}{2}
g_{AB}\nabla_{C}\Phi\nabla^{C}\Phi
\end{equation}
where only the kinetic terms are present; as it is customary, such a tensor can be derived from
a variational principle starting from a Lagrangian ${\cal L}_{\Phi}$ related to the scalar field $\Phi$, whose physical meaning will be clear below.

Because of the definition of 5D-space itself, it is important to stress now that no self-interaction potential $V(\Phi)$ has been taken into account so that $T_{AB}$ is a completely symmetric object and $\Phi$ is, by definition, a cyclic variable: this fact guarantees that Noether theorem holds for $T_{AB}$ and a conservation law intrinsically exists. With these considerations in mind, the field equations can now assume the form
\begin{equation}
\label{a11}
\mathcal{R}_{AB}=\chi\left(T_{AB}-\frac{1}{2}g_{AB}T\right)
\end{equation}
where $T$ is the trace of $T_{AB}$, $\chi$ is the 5D gravitational constant and $\hbar=c=1$. The form (\ref{a11}) of field equations is useful in order to put in evidence the role of the scalar field $\Phi$, if we are not simply assuming Ricci-flat 5D-spaces. 

As we said, $T_{AB}$ is a symmetric tensor, and due to the choice of the metric and to the symmetric nature of the stress-energy tensor $T_{AB}$ and of the Einstein field equations $G_{AB}$, the contracted Bianchi identities $\nabla_{A}T^{A}_{B}=0$ hold. Developing them gives
\begin{eqnarray}
\nonumber
&\nabla_{A}T^{A}_{B}=\nabla_{A}\left(\partial_{B}\Phi\partial^{A}\Phi-\frac{1}{2}
\delta_{B}^{A}\partial_{C}\Phi\partial^{C}\Phi
\right)=\\
\nonumber
&=\left(\nabla_{A}\Phi_{B}\right)\Phi^{A}+\Phi_{B}\left(\nabla_{A}\Phi^{A}\right)
-\frac{1}{2}\left(\nabla_{B}\Phi_{C}\right)\Phi^{C}
-\frac{1}{2}\Phi_{C}\left(\nabla_{B}\Phi^{C}\right)=\\
&=\left(\nabla_{A}\Phi_{B}\right)\Phi^{A}+\Phi_{B}\left(\nabla_{A}\Phi^{A}\right)-
\Phi_{C}\left(\nabla_{B}\Phi^{C}\right)
\label{ciccio1}
\end{eqnarray}
and since our 5D-space is a Riemannian manifold, then $\nabla_{A}\Phi_{B}=\nabla_{B}\Phi_{A}$ so that in this case, partial and covariant derivatives coincide for the scalar field $\Phi$. Finally we get
\begin{eqnarray}
\nabla_{A}T_{B}^{A}=\Phi_{B}\,^{(5)}\Box\Phi
\end{eqnarray}
where $^{(5)}\Box$ is the 5D d'Alembert operator defined as $\nabla_{A}\Phi^{A}\equiv g^{AB}\nabla_{A}\nabla_{B}\Phi\equiv \,^{(5)}\Box\Phi$. 

The general result is that the conservation of the energy tensor $T_{AB}$ implies the Klein-Gordon equation, which assigns the dynamics of $\Phi$, that is $\nabla_{A} T_{A}^{B}=0$ implies that $^{(5)}\Box \Phi=0$ assuming of course $\Phi_{B}\neq 0$ since we are dealing with a non-trivial physical field. As we shall see below, the conservation law of the 5D energy of the scalar field gives a physical meaning to the fifth dimension.

The reduction to the 4D-dynamics can be accomplished by taking into account the Campbell theorem \cite{campbell}: this theorem states that it is always possible to consider a 4D Riemannian manifold, defined by the line element $ds^2=g\dab dx\ua dx\ub$, in a 5D one with $dS^2=g_{AB}dx^A dx^B$. We have $g_{AB}=g_{AB}(x\ua,x^4)$ with $x^4$ the yet unspecified extra coordinate; as we discussed above, $g_{AB}$ is covariant under the group of 5D coordinate transformations but
not under the restricted group of 4D transformations. This fact has the relevant consequence that the choice of 5D coordinates results as the {\it gauge} necessary to specify the 4D physics also in its non-standard aspects. Viceversa, in specifying the 4D physics, the bijective embedding process in 5D gives physical meaning to the fifth coordinate $x^4$. 

In other words, the fifth coordinate, in 4D can assume a physical meaning, and we are going to see how this might be connected to the mass of elementary particles.

The reduction to the 4D-spacetime is encoded by the Lagrangian
\begin{eqnarray}
&{\cal L}=\sqrt{-g^{(5)}}\left[^{(5)}\mathcal{R}+\lambda(g_{44}-\epsilon\Phi^2)\right]
\end{eqnarray}
with $\lambda$ as a Lagrange multiplier, $\Phi$ a scalar field and $\epsilon=\pm 1$. This approach is completely general and used in theoretical physics when we want to put in evidence
some specific feature \cite{lagrange}; in this case, we need it to derive the physical gauge for the 5D-metric. 

We can write down the metric as
\begin{eqnarray}
&dS^2=g_{AB}dx^{A}dx^{B}= g\dab dx\ua dx\ub+g_{44}(dx^4)^2
=g\dab dx\ua dx\ub+\epsilon\Phi^2(dx^4)^2\,,
\label{a1"}
\end{eqnarray}
from which we obtain directly particle-like solutions for $\epsilon=-1$ or wave-like
solutions for $\epsilon=+1$ in the 4D-reduction procedure. The standard signature of 4D-component of the metric is $(+1,-1,-1,-1)$ and $\alpha,\beta=0,1,2,3$. 

Furthermore, the 5D-metric can be written in a Kaluza-Klein fashion as the matrix
\begin{equation}
\label{gmatrix}
 g_{AB}=\left(
\begin{array}{cc}
g\dab & 0\\ 0 & \epsilon\Phi^2
\end{array}
\right)\,,
\end{equation}
and the 5D-curvature Ricci tensor can be derived. After the projection from 5D to 4D, $g\dab$, derived from $g_{AB}$, no longer explicitly depends on $x^4$, and a useful expression for the Ricci scalar can be derived:
\begin{equation}
\label{ricci} ^{(5)}\mathcal{R}=\mathcal{R}-\frac{1}{\Phi}\Box \Phi\,,
\end{equation}
where the dependence on $\epsilon$ is explicitly disappeared and $\Box$ is the 4D-d'Alembert operator, which gives $\Box\Phi\equiv g\umunu\nabla_{\mu}\nabla_{\nu}\Phi$. 

The action above can therefore be written in a 4D-reduced Brans-Dicke action of the following form
\begin{equation}
{\cal L}=\frac{1}{16\pi G_{N}}\sqrt{-g}\left[\Phi \mathcal{R}+{\cal L}_{\Phi}\right]
\label{e1} 
\end{equation}
where the Newton constant $G_N$ is given in terms of the 5D gravitational constant and a characteristic length associated to the Compton length as $G_N=\frac{^{(5)}G}{2\pi l}$. Defining the function of the 4D-scalar field $\phi$ as
\begin{equation}
\label{pippo} 
-\frac{\Phi}{16\pi G_N}=\phi
\end{equation}
we get in 4D a general action in which gravity is non-minimally coupled to a scalar field as
\begin{eqnarray}
&\label{2.1.1}
{\cal A}=\int_{\cal M}d^{4}x\sqrt{-g}\left[\phi \mathcal{R}+\frac{1}{2}\gu\nabla_{\mu}\phi\nabla_{\nu}\phi-\v\right]+\int_{\pa {\cal M}}d^{3}x\sqrt{-b}K
\end{eqnarray}
where the form and the role of $\v$ are still general. The second integral is a boundary term where $K\equiv h^{ij}K_{ij}$ is the trace of the extrinsic curvature tensor $K_{ij}$ of the hypersurface
$\pa {\cal M}$ embedded in the 4D-manifold ${\cal M}$ and $b$ is the metric determinant of the 3D-manifold.

The Einstein field equations are derived by varying with respect to the 4D-metric $\gd$
\begin{eqnarray}
&\label{2.1.2} 
\mathcal{R}\dmunu-\frac{1}{2}\gd \mathcal{R}=T\dmunu
\end{eqnarray}
where
\begin{eqnarray}
&\label{2.1.4}
T\dmunu=\frac{1}{\phi}\left[-\frac{1}{2}\nabla_{\mu}\phi\nabla_{\nu}\phi
+\frac{1}{4}g\dmunu\nabla_{\alpha}\phi\nabla^{\alpha}\phi
-\frac{1}{2}g\dmunu\v-g\dmunu\Box\phi+\nabla_{\mu}\nabla_{\nu}\phi\right]
\end{eqnarray}
is the \emph{effective} energy tensor containing the non-minimal coupling contributions, the kinetic terms and the potential of the scalar field $\p$. 

By varying with respect to $\phi$, we get the 4D-Klein-Gordon equation, which are nothing else but the contracted Bianchi identity, as we have already shown; this implies that the effective energy tensor in the right-hand side of (\ref{2.1.2}) is a divergenceless tensor, which is compatible with Einstein theory of gravity also if we started from a 5D-space, as long as the theory is torsionless, that is in the vacuum of spinor fields.

Finally in order to physically identify the fifth dimension, we may write the Klein-Gordon equation in the form 
\begin{eqnarray}
&\label{kg1} \left(\Box + m_{eff}^2\right)\p=0
\end{eqnarray}
where
\begin{eqnarray}
&\label{kg2}m_{eff}^2=\frac{1}{\phi}\left[\vp-\mathcal{R}\right]
\end{eqnarray}
is the effective mass given in terms of scalar field self-interactions and external gravitational contributions. In any quantum field theory formulated on curved spacetimes, these contributions, at one-loop level, have the same weight, and indeed, Extended Theories of Gravity are renormalizable at one loop-level \cite{birrell}. 

What we want to show next is that a natural way to generate particle masses can be achieved starting from a 5D picture. In other words, the concept of mass can be derived from a geometric viewpoint.
\section{The generation of masses}
\subsection{Particle masses as eigenstates from 5D-spaces}
The above considerations allow a straightforward mechanism for the generation of masses that can be related to the projection from 5D to 4D-manifolds.

In a 5D-space, the 5D d'Alembert operator for particle-like solutions can be split selecting the
value $\epsilon=-1$ in the metric as $^{(5)}\Box=\Box-{\partial_4}^2$, so that for a given scalar field $\Phi$, we have
\begin{eqnarray}
&\label{embedding}
^{(5)}\Box\Phi=\left[\Box-{\partial_4}^2\right]\Phi=0
\end{eqnarray}
identically; because the (\ref{gmatrix} ) is diagonal in the fifth component, we may separate the variable as $\Phi=\p(t,\vec{x})\psi(x_4)$ and thus 
\begin{eqnarray}
&\frac{\Box\p}{\p}=\frac{1}{\psi}\left[\frac{d^2\psi}{dx_4^2}\right]=-k_n^2
\label{split2}
\end{eqnarray}
where $k_n$ must be a constant. From Eq.(\ref{split2}), we obtain the two equations of motion
\begin{eqnarray}
&\label{kg5} \left(\Box + k_n^2\right)\p=0\\
&\label{oscillation}\frac{d^2 \psi}{dx_4^2}+k_n^2\psi=0\,.
\end{eqnarray}
the former being the evolution equation for the scalar field with a mass term. 

The constant $k_n$ has the physical dimension of the inverse of a length and, assigning boundary conditions, we can derive the eigenvalue relation $k_n=\frac{2\pi}{l}n$ where $n$ is an integer. As a result, in standard units, we recover the Compton length $\lambda_n=\frac{\hbar}{2\pi m_n c}=\frac{1}{k_n}$ which assigns the mass of a particle. It has to be stressed that, the eigenvalues of Eq.(\ref{oscillation}) are the masses of particles which are generated by the reduction process from 5D to 4D. On the other hand, different values of $n$ fix the families of particles, while, for any given value of $n$.
\subsection{4D-spacetimes and dynamics of massive scalar fields}
Dynamics of 4D-component of the induced scalar field $\Phi$ can be related to the spacetime deformations discussed above; in this way, the role of $GL(4)$-group of diffeomorphism will result of fundamental importance. Let us start by showing how particles can acquire mass by deformations and let us relate the procedure with the above reduction mechanism. 

As usual, a particle with zero mass is characterized by the invariant relation $\eta^{\alpha\beta}p_{\alpha}p_{\beta}=0$ in the Minkowski spacetime. Deforming the spacetime and considering the above operators, one has
\begin{equation}\label{rm2}
\eta^{AB}\Phi_{A}^{C}\Phi_{B}^{D}e_{C}^{\alpha}e_{D}^{\beta}p_{\alpha}p_{\beta}=g^{\alpha\beta}p_{\alpha}p_{\beta}=0\,,
\end{equation}
so we have defined two frames, one, the Minkowski spacetime, defined by the metric $\eta^{\alpha\beta}$ and the other defined by the metric $g^{\alpha\beta}$, generated by the projection from 5D. The two frames are related by the matrices of deformation functions $\Phi_{A}^{C}(x)$. In both frames the massless particle follow a null path, but we observe that using the decomposition (\ref{controvariantdecomposition}) the particle does not appear massless with respect to the first frame. As matter of fact, Eq. (\ref{rm2}) becomes
\begin{equation}\label{rm3}
 \Omega^{-2}\eta^{\alpha\beta}p_{\alpha}p_{\beta}+\chi^{\alpha\beta}p_{\alpha}p_{\beta}=0\,,
\end{equation}
so that
\begin{equation}\label{rm33}
 \eta^{\alpha\beta}p_{\alpha}p_{\beta}=-\Omega^{2}\chi^{\alpha\beta}p_{\alpha}p_{\beta} \neq 0.
\end{equation}
and now in the first frame we are able to define a rest reference system for the particle.

For a massless particle it is not possible to define a rest reference system since considering the invariant relation,
\begin{equation}\label{rm4}
 g^{00}p^{2}_{0}+ 2g^{0i}p_{0}p_{i}+g^{ij}p_{i}p_{j}=0\,,
\end{equation}
and defining as rest frame the system in which $p_{j}=0$, the solution exists only for $p_{0}=0$ {\it i.e.} only the trivial solution $p_{\alpha}=0$ satisfies the `` rest reference frame'' condition.

On the other hand, if we consider massive particle, then
\begin{equation}\label{rm5}
 g^{00}p^{2}_{0}+ 2g^{0i}p_{0}p_{i}+g^{ij}p_{i}p_{j}=m^{2}\,,
\end{equation}
the rest frame is characterized by the conditions $p_{j}=0$ and consequently $g^{00}p^{2}_{0}=m^{2}$. This means that the (squared) mass is proportional to the (squared) energy and the solution is no more trivial. To overcome this problem, let us fix the deformation such that
\begin{equation}\label{rm6}
-\Omega^{2}\chi^{\alpha\beta}p_{\alpha}p_{\beta} = m^{2}\,,
\end{equation}
$m$ being the mass "attribute" to the particle. In the second frame we have
\begin{equation}\label{rm7}
p^{2}_{0}-{\vec{p}}^{\;2}+\Omega^{2}\chi^{00} p^{2}_{0} + 2\chi^{0i}p_{0}p_{i}+\chi^{ij}p_{i}p_{j}=0\,,
\end{equation}
if $\vec{p}=0$ then $p^{2}_{0}(1+\Omega^{2}\chi^{00})=1$ which implies, besides the trivial solution, also the condition $\Omega^{2}\chi^{00}=-1$ which, together with Eq. (\ref{rm6}), gives $p^{2}_{0}=m^{2}$, the rest system condition in the first frame.

It is also possible to show that there is an equivalence between deforming a metric and giving mass to a massless particle.
As we have shown, we have described the spacetime deformation by using matrices of scalar fields. The same problem can be
addressed in terms of spacetime tensors
\begin{equation}
\label{(36)}
\eta_{AB}\Phi^A_C\Phi^B_D e^C_\mu e^D_\nu =
 g_{\alpha\beta}\Phi^\alpha_\mu \Phi^\beta_\nu = \tilde g_{\mu\nu}\,.
\end{equation}
It should be observed that the $\Phi^\alpha_\mu$ do not represent coordinates transformations
as far as they cannot be reduced to Jacobian matrices. This means that $\Phi^\alpha_\mu$ have a fundamental physical meaning.
Starting from $\tilde g$ we observe that if $g_{\mu\nu}\,p^\mu\, p^\nu = 0$
then $\tilde g_{\mu\nu}\,p^\mu\, p^\nu \neq 0$ when $\tilde g$ is a general deformation that can be related to the conformal transformations of $g$.
Eq. (\ref{(36)}) tells us that, equivalently, we can read the deformation as a transformation
of the 4-momentum of the particle
\begin{equation}
\label{rm11}
\tilde g_{\mu\nu}\,p^\mu\, p^\nu =g_{\alpha\beta}\Phi^\alpha_\mu \Phi^\beta_\nu p^\mu\, p^\nu \equiv g_{\alpha\beta}\,\tilde p^\alpha\, \tilde p^\beta\,,
\end{equation}
in such a way there exists deformations of spacetime which give mass to massless particles. In other words, deformations, that are elements of the conformally-invariant $GL(4)$-group can be related, in principle, to the generation of the masses of particles.
\subsection{Mass Generation from Deformations}
So far we have considered the geometrical definition of mass from the point of view of relativistic mechanics. Now we would like to extend this result to classical field theory with the aim to extend it to quantum field theory.

Let us consider a scalar free massless particle. It can be described by the Lagrangian
\begin{eqnarray}
\label{la1}
{\cal L} &=& \frac{1}{2}\sqrt{-g}g^{\mu\nu}(\partial_\mu \phi )(\partial_\nu \phi )
=\frac{1}{2}\sqrt{-g}\left(\eta^{\mu\nu}+\chi^{\mu\nu}\right)(\partial_\mu \phi )(\partial_\nu \phi ).
\end{eqnarray}
It is well-known that the free propagator has a pole in $p^2=0$. In order to introduce a mass term in the Lagrangian and in the field equations, we have to eliminate a divergence from it, that is considering
\begin{eqnarray}\label{ft1}
&& \partial_{\mu}\left[ \partial_{\nu} \sqrt{-g}\,\chi^{\mu\nu}\phi^{2}\right]=\partial_{\mu}\partial_{\nu}\left(\sqrt{-g}\,\chi^{\mu\nu}\right)\phi^{2}
+\nonumber\\ &&+4\,\partial_{\nu}\left(\sqrt{-g}\,\chi^{\mu\nu}\right)\,\phi\,\partial_{\mu}\,\phi
+2\sqrt{-g}\,\chi^{\mu\nu}\partial_{\mu}\phi\;\partial_{\nu}\phi+2\;\sqrt{-g}\,\chi^{\mu\nu}\phi\;\partial_{\mu}\partial_{\nu}\phi\,,
\end{eqnarray}
the Lagrangian takes the form
\begin{eqnarray}\label{ft2}
\widetilde{{\cal L}} &=& \frac{1}{2}\sqrt{-g}\left(\eta^{\mu\nu} \right)(\partial_\mu \phi )(\partial_\nu \phi )- \frac{1}{4}\partial_{\mu}\partial_{\nu}\left(\sqrt{-g}\,\chi^{\mu\nu}\right)\phi^{2}-\nonumber\\ &&+ \,\partial_{\nu}\left(\sqrt{-g}\,\chi^{\mu\nu}\right)\,\phi\,(\partial_{\mu}\,\phi)-\frac{1}{2}\;\sqrt{-g}\,\chi^{\mu\nu}\phi\;\partial_{\mu}\partial_{\nu}\phi.\nonumber\\
\end{eqnarray}
In this way a ``mass'' term $m^{2}=\frac{1}{2}\partial_{\mu}\partial_{\nu}\left(\sqrt{-g}\,\chi^{\mu\nu}\right)$ can be defined in a new Lagrangian. On the other hand, considering an action ${\cal A}=\int\sqrt{- g}\widetilde{\mathcal{L}}$ and a variational principle $\frac{\delta {\cal A}}{\delta \phi}= 0$ implies the equation
\begin{equation}\label{ft5}
\frac{\partial \widetilde{\mathcal{L}}}{\partial \phi}-\partial_{\alpha}\frac{\partial \widetilde{\mathcal{L}}}{\partial (\partial_{\alpha}\phi)}+\partial_{\alpha}\partial_{\beta}\frac{\partial \widetilde{\mathcal{L}}}{\partial (\partial_{\alpha}\partial_{\beta}\phi)}=0\,,
 \end{equation}
which gives
\begin{equation}\label{ft7}
\Box\phi=\eta^{\mu\nu}\partial_{\mu}\partial_{\nu}\phi+\Omega^{2}\chi^{\mu\nu}\partial_{\mu}\partial_{\nu}\phi=0\,.
\end{equation}
It is well-known that this equation cannot give, in general, a potential or a mass term, except when we take as a solution a plane wave $\phi=\exp i k_{\mu} x^{\mu}$. In this case the $\chi$ part of the equation can be interpreted as a mass term according to
\begin{equation}\label{ft8}
\chi^{\mu\nu}\partial_{\mu}\partial_{\nu}\phi = -\Omega^{2}\chi^{\mu\nu}k_{\mu}k_{\nu} \exp i k_{\mu} x^{\mu}=m^{2}\phi\,.
\end{equation}
This equation can be compared with Eq. (\ref{rm6}), previously derived for a relativistic particle. We can extend this result to each function which can be expressed by a Fourier transform,
\begin{equation}\label{ft9}
\phi(x)=\int\exp (i k_{\mu} x^{\mu}) \tilde{f}(k)d^{4}k\,,
\end{equation}
where the spacetime dependent mass term is defined by the equation
\begin{equation}\label{ft10}
 m^{2}(x)=-\int \chi^{\mu\nu}k_{\mu}k_{\nu}\exp (i k_{\mu} x^{\mu}) \tilde{f}(k)d^{4}k.
\end{equation}
With these restrictions, the Klein-Gordon equation for a massless particle in the $g$-frame is seen in the $\eta$-frame as a {\it massive particle} as soon as the plane wave solutions are considered. This result makes more sense when the mass is interpreted as a quantum effect. In the case a curved spacetime, the above arguments can be implemented by the substitutions of operators $\eta \to g$ and $\partial\to\nabla$. It is important to note that the scalar field equation is not conformally invariant. In order to have a conformally invariant scalar field equation, it is necessary to introduce, in the Lagrangian density a non-minimal coupling between geometry and field. A possible choice is
\begin{equation}
\label{nmc1}
{\cal L} = \frac{1}{2}\sqrt{-g}g^{\mu\nu}(\partial_\mu \phi )(\partial_\nu \phi )
-\xi \mathcal{R}\phi^{2}\,,
\end{equation}
By varying with respect to $\phi$, the Klein-Gordon equation
\begin{equation}\label{nmc2}
 \Box \phi + \xi \mathcal{R} \phi=0\,,
\end{equation}
is recovered.
Minimal coupling is obtained for $\xi=0$. By analogy with the previous considerations, a deformation with constant curvature $\mathcal{R}=m^{2}$ gives a mass term in the original frame. However, we need also to interpret the other terms appearing in the new frame. Expanding Eq. (\ref{nmc2}),
\begin{equation}\label{nmc4}
\eta^{\mu\nu} \partial_{\mu}\partial_{\nu} \phi + \chi^{\mu\nu} \partial_{\mu}\partial_{\nu} \phi +\left(\eta^{\mu\nu}+\chi^{\mu\nu}\right)\Gamma_{\mu\nu}^{\lambda}\partial_{\lambda} \phi +\xi \mathcal{R} \phi=0\,,
\end{equation}
we need that the connection is compatible with the metric tensor $(\eta^{\mu\nu}+\chi^{\mu\nu}) $, that is
\begin{equation}\label{nmc5}
\nabla \left(\eta^{\mu\nu}+\chi^{\mu\nu}\right)=0.
\end{equation}
In order to have a deformation defining a mass $m^{2}=\mathcal{R}$ in the original frame, two conditions must be satisfied, $\mathcal{R}=\mathcal{R}_0 >0$ where $\mathcal{R}_0$ is a constant and
\begin{equation}
\label{nmc66}
\chi^{\mu\nu} \partial_{\mu}\partial_{\nu} \phi +
\left(\eta^{\mu\nu}+\chi^{\mu\nu}\right)\Gamma_{\mu\nu}^{\lambda}\partial_{\lambda} \phi=0\,,
\end{equation}
which is a restriction on the deformation tensor $\chi^{\mu\nu}$. This condition allows to determine the mass by a conformal transformation. In fact, in the new frame, the equation is%
\begin{equation}\label{nmc7}
 g^{\alpha\beta}\nabla_{\alpha}\partial_{\beta}\phi+\xi \mathcal{R} \phi=0\,,
\end{equation}
which, in the case of a conformal transformation $g_{\alpha\beta}=\Omega^{2}\eta_{\alpha\beta}$ becomes
\begin{equation}\label{nmc9}
\Omega^{-2}\eta^{\alpha\beta}\left(\partial_{\alpha}\partial_{\beta}\phi + 2 \Omega^{-1}\partial^{\gamma}\Omega\partial_{\gamma}\phi\right)-6\xi\eta^{\alpha\beta}\Omega^{-3}\partial_{\alpha}\partial_{\beta}\Omega=0\,.
\end{equation}
It is equivalent to a massive scalar field equation (in the first frame), if the condition
\begin{equation}\label{nmc10}
 2 \Omega^{-1}\partial^{\gamma}\Omega\partial_{\gamma}\phi -6\xi\eta^{\alpha\beta}\Omega^{-1}\partial_{\alpha}\partial_{\beta}\Omega=m^{2}\,,
\end{equation}
holds. Since Eq. (\ref{nmc9}) is linear in $\phi$, we can define $\phi=e^{i\lambda_{\alpha}x^{{\alpha}}}$. Eq. (\ref{nmc10}) becomes
\begin{equation}\label{nmc11}
 2 \Omega^{-1}\partial^{\gamma}\Omega \lambda_{\gamma} -6\xi\eta^{\alpha\beta}\Omega^{-1}\partial_{\alpha}\partial_{\beta}\Omega=m^{2}\,,
\end{equation}
then a solution is $ \Omega=e^{ik_{\alpha}x^{{\alpha}}} $ and the mass depends on the combination
\begin{equation}\label{nmc12}
 m^{2}=6\xi \eta^{\alpha\beta}k_{\alpha}k_{\beta}-2\eta^{\alpha\beta}k_{\alpha}\lambda_{\beta}.
\end{equation}
In this example, the deformation is a specific complex function, but we can extend this result to any function by taking
\begin{equation}\label{nmc13}
\Omega(x)=\int \tilde{\Omega}(k)e^{ikx}d^{4}k\,,
\end{equation}
and also non-constant masses can be is obtained being
\begin{equation}\label{nmc14}
m^{2}=\frac{\int \tilde{\Omega}(k)(6\xi \eta^{\alpha\beta}k_{\alpha}k_{\beta}-2\eta^{\alpha\beta}k_{\alpha}\lambda_{\beta})e^{ikx}d^{4}k}{\int \tilde{\Omega}(k)e^{ikx}d^{4}k}\,.
\end{equation}
These results mean that a geometrical definition of mass is always possible. It can be induced by the conformal transformation $\Omega$ a restriction of the deformations group $GL(4)$. 

Summarizing we have shown that the fifth dimension of a reduction mechanism from 5D to 4D-manifolds can be interpreted as a mass generator, where the degrees of freedom coming from ETGs have a physical meaning and cannot be simply gauged away, and as soon as particles acquire masses $GL(4)$ can induce symmetry breaking.
In other words, the further degrees of freedom related to the ETGs may be related to a Higgs-like mechanism as we are going to discuss below.
\section{Gravitational massive states and induced symmetry breaking}
The above results could be interesting to investigate quantum gravity effects and symmetry breaking in the range between GeV and TeV scales. Such scales are actually investigated by the experiments at LHC. It is important to stress that any ultra-violet model of gravity (e.g. at TeV scales) have to explain also the observed weakness of gravitational effects at largest (infra-red) scales. This means that massless modes have to be considered in any case.

The above 5D-action is an example of higher dimensional action where the effective gravitational energy scale (Planck scale) can be rescaled according to Eq. (\ref{pippo}). In terms of mass, being $M_p^2=\frac{c\hbar}{G_N}$ the constraint coming from the ultra-violet limit of the theory ($10^{19}$ GeV) , we can set $M_p^2 = M_{cut-off}^{D-2} V_{D-4}$, where $V_{D-4}$ is the volume coming from the extra dimension. It is easy to see that $V_{D-4}$, in the 5D case, is related to the fifth component of $\Phi$. $M_{cut-off}$ is the cut-off mass that becomes relevant as soon as the Lorentz invariance is violated. Such a scale, in the context discussed here, could be of the order TeV \cite{calmet}.

As we have shown above, it is quite natural to obtain effective theories containing scalar fields of gravitational origin. In this sense, $M_{cut-off}$ is the result of dimensional reduction. To be more explicit, the 4D dynamics is led by the effective potential $V(\phi)$ and the non-minimal coupling $F(\phi)$. Such functions could be tested since they are related to massive states.

In particular, the effective Extended Gravity, produced by the reduction mechanism from 5D to 4D, can be chosen as 
\begin{eqnarray}
{\cal A}=\int d^4x ~\sqrt{-g} ~\Big [ -\frac{\phi^2}{2} \mathcal{R}
~+\frac{1}{2} g^{\mu\nu}\partial_\mu\phi \partial_\nu\phi
-V\Big]
\label{model}
\end{eqnarray}
plus contributions of ordinary matter terms. The potential for $\phi$ can be assumed as
\begin{eqnarray}
V(\phi) = \frac{M_{cut-off}^2}{2} \phi^2 + \frac{\lambda}{4} \phi^4\,,
\end{eqnarray}
where a massive term and the self-interaction term are present. This is the standard choice of quantum field theory which perfectly fits with the arguments of dimensional reduction. Let us recall again that the scalar field $\phi$ is not put {\it by hand} into dynamics but it is given by the extra degrees of freedom of gravitational field generated by the reduction process in 4D. It is easy to derive the vacuum expectation value of $\phi$, being $M_{cut-off}^2 = 2\lambda M_{p}^2$, which is a fundamental scale of the theory.

Some considerations are in order at this point. Such a scale has to be confronted with Higgs vacuum expectation value which is 246 GeV and then with the {\it hierarchy problem}. If $M_{cut-off}$ is larger than Higgs mass, the problem is obviously circumvented. It is important to stress that hierarchy problem occurs when couplings and masses of effective theories are very different than the parameters measured by experiments. This happen since measured parameters are related to the fundamental parameters by renormalization. However cancellations between fundamental quantities and quantum corrections are necessary, then the hierarchy problem is a fine-tuning problem. 

In particle physics, the puzzle is why the weak force is stronger and stronger than gravity. Both of these forces involve fundamental constants of nature, the Fermi constant for the weak force and the Newton constant for gravity. From the Standard Model of Particles, the Fermi constant is unnaturally large and should be of the order of the Newton constant, unless there is a cancellation between the bare value of Fermi constant and the quantum corrections to it. 

More technically, the question is why the Higgs boson is lighter and lighter than the Planck mass. In fact, people are searching for Higgs masses ranging from 115 up to 350 GeV with different selected decay channels from $b\bar{b}$ to $t\bar{t}$ (see for example \cite{camy} and references therein). One would expect that the large quantum contributions to the square of the Higgs boson mass would inevitably make the mass huge, comparable to the scale at which new physics appears, unless there is a fine-tuning cancellation between the quadratic radiative corrections and the bare mass. With this state of art, the problem cannot be formulated in the context of the Standard Model where the Higgs mass cannot be calculated. In a sense, the problem is solvable if, in a given effective theory of particles, where the Higgs boson mass is calculable, there are no fine-tunings. If one accepts the {\it big-desert} assumption and the existence of a hierarchy problem, some new mechanism (at Higgs scale) becomes necessary to avoid fine-tunings.

The model which we are discussing contains a "running" scale that could avoid to set the Higgs scale with great accuracy. If the mass of the field $\phi$ is in TeV region, there is no hierarchy problem being $\phi$ a gravitational scale. In this case, the Standard Model holds up plus an extended gravitational sector that could be derived from the fifth dimension.
In other words, the Planck scale can be dynamically derived from the vacuum expectation value of $\phi$. 
In our model, the Planck scale can be recovered, as soon as the coupling $\lambda$ is of order $10^{-30\div31}$. 

Considering again the problem of mass generation, one can assume that particles of Standard Model have sizes related to the cut off, that is $M_{cut-off}^{-1}$, and their collisions could lead to the formation of bound states as in \cite{Antoniadis:2001sw, Antoniadis:1998ig}. Potentially, such a phenomenon could mimic the decay of semi-classical quantum black holes and, at lower energies, it could be useful to investigate substructures. This means that we should expect some strong scattering effects in the TeV region involving the coupling of $\phi$ to the Standard Model fields. The "signature" of this phenomenon could lead to polarization effects of the particle beam as discussed in \cite{sabafe}. Furthermore the strong dynamics derived from the phenomenon could resemble compositeness as discussed in \cite{Meade:2007sz}. 
These considerations are extremely important in view of Higgs sector physics.

In fact the Higgs mechanism is an approach that allows to generate the masses of electroweak gauge bosons; to preserve the perturbative unitarity of the S-matrix; to preserve the renormalizability of the theory. The masses of the electroweak bosons can be written in a gauge invariant form using either the non-linear sigma model \cite{CCWZ2} or a gauge invariant formulation of the electroweak bosons. However if there is no propagating Higgs boson, quantum field amplitudes describing modes of the electroweak bosons grow too fast violating the unitarity around TeV scales \cite{LlewellynSmith:1973ey,Lee:1977yc,Lee:1977eg,Vayonakis:1976vz}. There are several ways in which unitarity could be restored but the Standard Model without a Higgs boson results non-renormalizable. 

A possibility is that the weak interactions become strongly coupled at TeV scales and then the related gauge theory becomes unitary at non-perturbative level. Another possibility for models without a Higgs boson consists in introducing weakly coupled new particles to delay the unitarity problem into the multi TeV regime where the UV limit of the Standard Model is expected to become relevant. In \cite{Dvali:2010jz}, it is proposed that, as black holes in gravitational scattering, classical objects could form in the scattering of W-bosons. 

This idea shows several features of electroweak interactions. First of all, the Higgs mechanism is strictly necessary to generate masses for the electroweak bosons. Beside, some mechanisms can be unitary but not renormalizable or vice-versa. In summary, the paradigm is that three different criteria should be fulfilled: $i)$ a gauge invariant generation of masses of electroweak bosons, $ii)$ perturbative unitarity; $iii)$ renormalizability.
 The approach we are proposing here is based on Extended Theories of Gravity deduced from a 5D-manifold reduction where the Standard Model is fully recovered enlarging the gravitational sector and avoiding the hierarchy problem.

It is important to point out that, in both the non-linear sigma model and in gauge invariant formulation of Standard Model, it is possible to define an action in terms of an expansion in the scale of the electroweak interactions $v$. The action can be written as 
\begin{eqnarray} \label{effaction}
{\cal A}={\cal A}_{SM w/o Higgs}+\int d^4 x \sum_i \frac{C_i}{v^N} O^{4+N}_i\,,
\end{eqnarray}
where $O^{4+N}_i$ are operators compatible with the symmetries of the model. 
The analogy between the effective action for the electroweak interactions (\ref{effaction}) and that of Extended Gravity is striking. Considering only the leading terms, the above theory can be written as a Taylor series of the form 
\begin{equation}
\label{effaction2}
f(R)\simeq \Lambda+\sum_k \frac{1}{k!}\left.\frac{d^{k}f(R)}{dR^{k}}\right|_{0} R^{k}
\end{equation}
where the coefficients are the derivatives of $f(R)$ calculated at a certain value of $R$. Clearly, as shown in previous sections,, the extra gravitational degrees of freedom can be suitably transformed in a scalar field $\phi$ which allows to avoid the hierarchy problem. Both electroweak theory and Extended Gravity have a dimensional energy scale which defines the strength of the interactions. The Planck mass sets the strength of gravitational interactions while the weak scale $\lambda$ determines the range and the strength of the electroweak interactions. As shown in the previous subsection, these scales can be compared at TeV energies. 

In other words, the electroweak bosons are not gauge bosons in standard sense but they can be "derived" from the above further gravitational degrees of freedom. The local $SU(2)_L$ gauge symmetry is imposed at the level of the quantum fields. However there is a residual global $SU(2)$ symmetry, {\it i.e.} the custodial symmetry. In the case of gravitational theories formulated as the $GL(4)$-group of diffeomorphisms, tetrads are an unavoidable feature necessary to construct the theory. They are gauge fields which transform under the local Lorentz transformations $SO(3,1)$ and under general coordinate transformations, the metric $g_{\mu\nu}= e^a_\mu e^b_\nu \eta_{ab}$ which is the field that is being quantized, transforms under general coordinate transformations which is the equivalent of the global $SU(2)$ symmetry for the weak interactions (in our case the residual $GL(2)\supset SU(2)$) . Such an analogy between the tetrad fields and the Higgs field is extremely relevant. As shown above for deformations, we can say that the Higgs field has the same role of the tetrads for the electroweak interactions while the electroweak bosons have the same role of the metric. Dynamics would be given by deformations. 

A gravitational action like (\ref{effaction2}) is, in principle, non-perturbatively renormalizable if, as shown by Weinberg, there is a non-trivial fixed point which makes the gravity asymptotically free \cite{fixedpoint}. This scenario implies that only a finite number of the Wilson coefficients in the effective action would need to be measured and the theory would thus be predictive and probed at LHC. 

In conclusion, the unitarity problem of the weak interactions could be fixed by a non-trivial fixed point in the renormalization group of the weak scale. A similar mechanism could also fix the unitarity problem for fermions masses \cite{Appelquist:1987cf, Maltoni:2000iq, Maltoni:2001dc, Dicus:2004rg, Dicus:2005ku} if their masses are not generated by the standard Higgs mechanism but in the same way considered here (let us remind that also $SU(3)$ could be generated by the splitting of $GL(4)$ group). In the case of electroweak interactions this approach could be soon checked at LHC but good indications are also available for QCD \cite{atlas}.
\section{$f(R)$-gravity with torsion and spinor fields}
Considering torsion in the discussion means to take into account also spin fields \cite{Hehl2}.
In particular, matter fields can be given by Dirac fields: Dirac fermions are defined in terms of gamma matrices satisfying the Clifford algebra. Such fields can be achieved 
in the process of 4D-reduction considering the whole geometric budget where generic connections are assumed (not only symmetric ones). The fifth gamma matrix is a parity-odd gamma matrix whose index has no geometrical meaning: it is $\gamma^{5}=i\gamma^{0}\gamma^{1}\gamma^{2}\gamma^{3}$ in terms of the other independent gamma matrices. 
The dynamics of the gravitational field can be taken in a more general instance: that is instead of considering the usual Hilbert-Einstein Lagrangian $R$, we will take a general function of the Hilbert-Einstein Lagrangian $f(R)$ derived from the reduction process \cite{sabafe}. The special case given by the Einstein-Hilbert Lagrangian is immediately recovered for $f(R)=R$, but in this case the peculiar form of the Lagrangian is such that yields an Hessian determinant null, and therefore it is degenerated with respect to all other more general $f(R)$-theories. In the following we will focus on these most general $f(R)$-theories of gravitation which can be always conformally related to the presence of scalar fields \cite{book,physrep}.
Dirac fermions possess in general both energy and spin density, and so their coupling to both metric and torsion is very natural: for them, it is possible to write an effective Lagrangian as
\begin{eqnarray}
&{\cal L}=\sqrt{-g}\left[f(R)
+\frac{i}{2}\left(\overline{\psi}\gamma^{\alpha}D_{\alpha}\psi
-\overline{\psi}\gamma^{\alpha}D_{\alpha}\psi\right)-m\overline{\psi}\psi\right]
\end{eqnarray}
that is the Dirac action plus a gravitational action given by a general function $f(R)$ of the Hilbert-Einstein action of gravity. Note that now we are including torsion degrees of freedom. By varying this action with respect to the fields involved, we get the field equations in the form
\begin{eqnarray}
&i\gamma^{\alpha}D_{\alpha}\psi+i\gamma^{\alpha}T_{\alpha}\psi-m\psi=0
\end{eqnarray}
where $m$ is the mass of the Dirac matter field and $T_{\alpha}$ is the contorsion vector (see Section 2). Energy and spin density are
\begin{subequations}
\label{consquantities}
\begin{eqnarray}
\label{energy}
&T_{\alpha\beta}=\frac{i}{4}\left(\bar{\psi}\gamma_{\alpha}D_{\beta}\psi-D_{\beta}\bar{\psi}\gamma_{\alpha}\psi\right)\\
\label{spin}
&S_{\mu\alpha\beta}=\frac{i}{4}\bar{\psi}\left\{\gamma_{\mu},[\gamma_{\alpha},\gamma_{\beta}]\right\}\psi
\end{eqnarray}
\end{subequations}
the spin density tensor being completely antisymmetric; the field equations for gravity are
\begin{subequations}
\label{geometricequations}
\begin{eqnarray}
\label{curvature-energy}
&f'(R)R_{\alpha\beta}-\frac{1}{2}f(R)g_{\alpha\beta}=T_{\alpha\beta}\\
\label{torsion-spin}
&f'(R)T_{\mu\alpha\beta}-\frac{1}{4}\left(g_{\alpha\mu}\partial_{\beta}f'(R)-g_{\beta\mu}\partial_{\alpha}f'(R)\right)
=\frac{1}{2}\left(S_{\alpha\beta\mu}+\frac{1}{2}g_{\alpha\mu}S_{\beta \rho}^{\;\;\;\rho}-\frac{1}{2}g_{\beta\mu}S_{\alpha\rho}^{\;\;\;\rho}\right)
\end{eqnarray}
\end{subequations}
After substitution of energy and spin density into these field equations, we get
\begin{subequations}
\label{geomfieldequations}
\begin{eqnarray}
\label{curv-matterenergy}
&f'(R)R_{\alpha\beta} -\frac{1}{2}f(R)g_{\alpha\beta}
=\frac{i}{4}\left(\bar{\psi}\gamma_{\alpha}D_{\beta}\psi-D_{\beta}\bar{\psi}\gamma_{\alpha}\psi\right)\\
\label{tors-matterspin}
&f'(R)T_{\mu\alpha\beta}-\frac{1}{4}\left(g_{\alpha\mu}\partial_{\beta}f'(R)-g_{\beta\mu}\partial_{\alpha}f'(R)\right)
=\frac{i}{8}\bar{\psi}\left\{\gamma_{\mu},[\gamma_{\alpha},\gamma_{\beta}]\right\}\psi
\end{eqnarray}
\end{subequations}
linking the Dirac matter to the spacetime geometry. Here $f'(R)$ means the ordinary derivative with respect to $R$. We see that when the Dirac matter field equations are used for the energy and spin density, the geometric field equations reduce to the Bianchi identities, showing that the model is consistently defined. It is important to stress that we are again in the realm of ETGs and that $f(R)$-gravity is nothing else but a scalar-tensor gravity where the 4D-scalar field $\phi$ has been suitably recast in terms of the curvature scalar $R$ [specifically $\phi\rightarrow f'(R)]$ \cite{noitor}.

It is possible to see that from \eqref{torsion-spin} there could be torsion even in absence of the spin density tensor: therefore we may say that while the energy is the source of the curvature of the spacetime, both the spin and the non-linearity of $f(R)$ gravity are the sources of the spacetime torsion, in particular the non-linearity being the source of the torsion trace vector; however in the case of Dirac field, because of the complete antisymmetry of the spin, matter is the source of the completely antisymmetric part of torsion alone.

In the following, we suppose that the trace of the field energy-curvature coupling equation \eqref{curvature-energy} given by $f'(R)R-2f(R)=T^{\alpha}_{\alpha}=T$ is an invertible relation between the Ricci curvature scalar $R$ and the trace of the stress-energy tensor; since $f(R)=kR^2$ is only compatible with the traceless energy condition, then we assume that $f(R)\not=kR^2$.

Clearly, it is now possible to split all torsional quantities into torsionless quantities plus torsional contributions. The Dirac theory is algebraically related to the spin, therefore we can substitute torsion with the spin everywhere, to get the Dirac field equations
\begin{eqnarray}
\label{Dirac}
&i\gamma^{\alpha}\nabla_{\alpha}\psi
-\frac{3}{16\varphi}\bar{\psi}\gamma^{\alpha}\psi\gamma_{\alpha}\psi-m\psi=0
\end{eqnarray}
where $\nabla$ designates the purely metric derivatives and the contributions of torsion are written in terms of the spin of the spinor field itself. Here have introduced the field $\varphi=f'(R)$: this means that Dirac field equations\footnote{Here we are using the symbol $\varphi$ instead of $\phi$ since torsional degrees of freedom are present in $f(R)$.} , in the most general case with torsion, are dynamically equivalent to those with no torsion but with additional self-interaction potentials (see consideration in the previous sections). By employing the Fierz rearrangements they can be written in the form of Nambu--Jona--Lasinio potentials; moreover this means that Dirac field equations in the most general $f(R)$-gravity are dynamically equivalent to those in the simplest case $f(R)=R$ in which the additional potentials of self-interaction are scaled by a factor $\varphi$, which may be either positive or negative yielding repulsive or attractive Nambu-Jona--Lasinio potentials respectively \cite{n-j--l,f-v}.

In the 4D case, and in absence of spinor fields, it is possible to work out the field equations that we would obtain by varying with respect to the metric only, as in the purely metric case, that is 
\begin{equation}
\label{h2}
f'(\mathcal{R})\mathcal{R}\dab-\frac{1}{2}f(\mathcal{R})g\dab
=\left(g_{\alpha\mu}g_{\beta\nu}-g\dab\gd\right)\nabla^{\mu}\nabla^{\nu}f'(\mathcal{R})\,, 
\end{equation} 
which are fourth-order equations; after a suitable manipulation, the above equation can be rewritten as
\begin{eqnarray}
\label{h4}
\mathcal{R}\dab-\frac{1}{2}\mathcal{R}g\dab
=\frac{1}{\phi}\left[\frac{1}{2}g\dab\left(f(\mathcal{R})-\mathcal{R}f'(\mathcal{R})\right)
+\nabla_{\alpha}\nabla_{\beta}f'(\mathcal{R})+g\dab\Box f'(\mathcal{R})\right]
\end{eqnarray}
where the right-hand side can be interpreted as an effective energy tensor constructed by the extra curvature terms \cite{book,physrep}.
\subsection{Torsion and weak interactions}
Having established a general framework in which the gravitational background can be related to the electroweak phenomenology, we shall now focus on a specific example in which such an approach can be realized.
It is easy to see that the field $\frac{1}{\varphi}$ plays the role of an energy-dependent scaling. Let us assume that the action \eqref{effaction2} is truncated at the fourth-order power: the Taylor coefficients can be written as
\begin{eqnarray}
\label{formf}
&f(R)=R+\frac{1}{4}\varepsilon R^{2}+\frac{1}{27}\eta^{2}\varepsilon^{2}R^{3}
\end{eqnarray}
in terms of the two parameters $\varepsilon$ and $\eta$ that have to be determined. The $\Lambda$ term has been discarded since it is unessential for the following considerations.
In this situation, the trace of the energy-curvature coupling is a cubic equation in $R$ with three solutions, of which we are going to take into account the only real one; this is also the only solution for which $R$ vanishes as the trace of the energy-momentum tensor $\Sigma$ vanishes. In this sense, this is the correct infrared limit. Explicitly $\varphi$ is
\begin{eqnarray}
\label{formvarphi}
&\varphi=3\left(1-\frac{3}{8\eta^{2}}\right)
+\left(\frac{3}{4\eta}+\sqrt[3]{s+\sqrt[2]{s^{2}-1}}\right)^{2}
+\left(\frac{3}{4\eta}+\sqrt[3]{s-\sqrt[2]{s^{2}-1}}\right)^{2}
\end{eqnarray}
function of the energy-momentum trace $\frac{1}{2}\eta\varepsilon\Sigma=s$ in the only parameter $\eta$.
This $\frac{1}{\varphi}$ energy-dependent coupling starts from the unity in the infrared regime, and then it increases up to its maximum value reached at some scale that may be fine-tuned eventually, before decreasing to vanish asymptotically in the ultraviolet regime. Henceforth any Nambu-Jona--Lasinio interaction will be negligible in the infrared, then it would get larger up to a given energy, before ensuring asymptotic freedom in the ultraviolet regime.

Having established how the non-linearity of $f(R)$-gravity behaves as an energy-dependent running coupling for a given Nambu-Jona--Lasinio interaction, we may now leave the treatment of the single-field in self-interaction to study the couple of fields in interaction. In this case, the role of torsion is drastically important for the following reason: each spinor has a dynamics governed by a covariant derivative that contains torsion related to the spin density of the total system.
If we now have only two spinor fields in a region of spacetime getting close to one another, it is clear that either of them will have dynamical behavior influenced by the presence of the other one, according to whether the other spinor is actually close enough to make relevant contributions to the spin density or far enough to consider those contributions to the spin density negligible. This means that the presence of torsion for a system of two or many spinors has effects for their dynamics, which make them look like interacting even if no other interaction has been defined.

The presence of the running coupling may fine-tune the scale of the interaction to any value. Consequently it is straightforward to wonder if this torsionally induces interaction might be similar to the nuclear interactions, somehow. These torsional potentials are similar to the forces in the Fermi model: this hypothesis was conjectured since the 1970s in some seminal works \cite{Hehl1,Hehl2,s-s,s-g,s-s/2,s-s/1} although one of the problems people had to face was precisely that of the energy-dependent scale: since this problem appears to be addressable by simply setting the non-linearity of the present gravitational action, we may ask ourselves what are the possible effects of torsional interactions and, in particular, if they may look like the Fermi nuclear interactions.
\subsection{The massless spinor-semispinor system} 
Let us consider now massless fields, so to split the left-handed and right-handed projections. As a first example, we shall deal with the case in which one of the two spinors is a semispinor having only the left-handed projection \cite{f/1}. The aim is to compare this case with that of the electroweak interactions for leptons before the symmetry breaking.
In the case of massless spinors of which one is a spinor and the other is a semispinor (having only the left-handed projection) we set $\psi^{1}_{R}\equiv0$.
The field equations are
\begin{eqnarray}
&i\gamma^{\mu}\nabla_{\mu}\psi^{1}_{L}
+\frac{3}{16}\left(\bar{\psi}^{1}_{L}\gamma_{\mu}\psi^{1}_{L}\gamma^{\mu}\psi^{1}_{L}
+\bar{\psi}^{2}_{L}\gamma_{\mu}\psi^{2}_{L}\gamma^{\mu}\psi^{1}_{L}
-\bar{\psi}^{2}_{R}\gamma_{\mu}\psi^{2}_{R}\gamma^{\mu}\psi^{1}_{L}\right)=0\\
&i\gamma^{\mu}\nabla_{\mu}\psi^{2}_{L}
+\frac{3}{16}\left(\bar{\psi}^{1}_{L}\gamma_{\mu}\psi^{1}_{L}\gamma^{\mu}\psi^{2}_{L}
+\bar{\psi}^{2}_{L}\gamma_{\mu}\psi^{2}_{L}\gamma^{\mu}\psi^{2}_{L}
-\bar{\psi}^{2}_{R}\gamma_{\mu}\psi^{2}_{R}\gamma^{\mu}\psi^{2}_{L}\right)=0\\
&i\gamma^{\mu}\nabla_{\mu}\psi^{2}_{R}
-\frac{3}{16}\left(\bar{\psi}^{1}_{L}\gamma_{\mu}\psi^{1}_{L}\gamma^{\mu}\psi^{2}_{R}
+\bar{\psi}^{2}_{L}\gamma_{\mu}\psi^{2}_{L}\gamma^{\mu}\psi^{2}_{R}
-\bar{\psi}^{2}_{R}\gamma_{\mu}\psi^{2}_{R}\gamma^{\mu}\psi^{2}_{R}\right)=0
\end{eqnarray}
which can be written in the form
\begin{eqnarray}
&i\gamma^{\mu}\nabla_{\mu} L
-\frac{1}{2}g\vec{A}_{\mu}\cdot\vec{\sigma}\gamma^{\mu}L
+\frac{1}{2}g'B_{\mu}\gamma^{\mu}L-G_{Y}\phi R=0
\label{Leftlepton}\\
&i\gamma^{\mu}\nabla_{\mu} R+g'B_{\mu}\gamma^{\mu}R-G_{Y}\phi^{\dagger}L=0
\label{Rightlepton}
\end{eqnarray}
by defining
\begin{eqnarray}
&\left(\psi^{2}_{R}\right)=R\ \ \ \ \ \ \ \ 
\left(\begin{tabular}{c}
$\psi^{1}_{L}$\\ $\psi^{2}_{L}$
\end{tabular}\right)=L
\end{eqnarray}
with
\begin{eqnarray}
&\frac{3}{8}\bar{R}L=\phi
\end{eqnarray}
and
\begin{eqnarray}
&\frac{3}{8}\left(\bar{L}\gamma_{\mu}\frac{\mathbb{I}}{2}L+\bar{R}\gamma_{\mu}R\right)
=-\left(\frac{g'}{1-G_{Y}}\right)B_{\mu}\\
&\frac{3}{8}\bar{L}\gamma_{\mu}\frac{\vec{\sigma}}{2}L=\left(\frac{3g}{3-G_{Y}}\right)\vec{A}_{\mu}
\end{eqnarray}
as new fields. We see that field equations (\ref{Leftlepton}) and (\ref{Rightlepton}) are formally identical to the field equations for the lepton fields before the symmetry breaking in the Standard Model.

We notice that from the fundamental lepton fields, it is possible to build up a complex singlet and a complex doublet: from these, we also see that the composite scalar field is a complex doublet; and the composite vector fields are a real singlet and a real triplet. In fact, consider that the massless fundamental leptons $R$ and $L$ are complex and therefore they can be subject to two independent $U(1)$ transformations: the massless composite scalar field, given by the definition $\frac{3}{8}\bar{R}L$, is consequently complex and thus it transforms according to the combined $U(1)$ transformations. Finally the massless composite vector fields are real and do not transform at all. Moreover we have that the massless fundamental leptons $R$ and $L$ are such that $R$ is the only right-handed spinor and it does not transform into anything else whereas $L$ is formed by a doublet of left-handed spinors which, in principle, are indistinguishable and therefore they can transform into one another according to an $SU(2)_{L}$ transformation: the massless composite scalar field $\frac{3}{8}\bar{R}L$ transforms according to the $SU(2)_{L}$ transformation above; furthermore, the massless composite vector fields are such that the vector field, given by the right-handed projections $\bar{R}\gamma^{\mu}R$, does not transform whereas, on the other hand, the vector fields, given by the left-handed projections, are such that the vector field, given by $\bar{L}\gamma_{\mu}L$, transforms into itself. The three vector fields $\bar{L}\gamma_{\mu}\vec{\sigma}L$ transform into each other and so the vector fields given by $2\bar{R}\gamma^{\mu}R+\bar{L}\gamma_{\mu}L$ and $\bar{L}\gamma_{\mu}\vec{\sigma}L$ transform as the singlet and the triplet of the adjoint representation of the same $SU(2)_{L}$ transformation that has been given above. This establishes the transformation laws of the lepton, the scalar and vector fields before the symmetry breaking in the Standard Model.
\subsection{The massless spinor-spinor system} 
As a second example, we shall deal with the case in which two spinors have both projections, but for which the two right-handed projections have different Abelian charges so to be forbidden to mix \cite{f/2}. The aim is to compare this case with that of the electroweak interactions for quarks before the symmetry breaking.

In the case of massless spinors having both projections, there is nothing that would forbid the two right-hand projections to mix forming a doublet, contrary to what is observed; to avoid this situation maintaining the two right-handed projections as two singlets, we have to postulate in addition that Abelian fields are introduced and the charges are assigned to the two right-handed fields with two different values. 
The field equations then are
\begin{eqnarray}
&i\gamma^{\mu}\nabla_{\mu}\psi^{1}_{L}
+\frac{3}{16}\left(\bar{\psi}^{1}_{L}\gamma_{\mu}\psi^{1}_{L}\gamma^{\mu}\psi^{1}_{L}
+\bar{\psi}^{2}_{L}\gamma_{\mu}\psi^{2}_{L}\gamma^{\mu}\psi^{1}_{L}
-\bar{\psi}^{1}_{R}\gamma_{\mu}\psi^{1}_{R}\gamma^{\mu}\psi^{1}_{L}
-\bar{\psi}^{2}_{R}\gamma_{\mu}\psi^{2}_{R}\gamma^{\mu}\psi^{1}_{L}\right)=0\\
&i\gamma^{\mu}\nabla_{\mu}\psi^{1}_{R}
-\frac{3}{16}\left(\bar{\psi}^{1}_{L}\gamma_{\mu}\psi^{1}_{L}\gamma^{\mu}\psi^{1}_{R}
+\bar{\psi}^{2}_{L}\gamma_{\mu}\psi^{2}_{L}\gamma^{\mu}\psi^{1}_{R}
-\bar{\psi}^{1}_{R}\gamma_{\mu}\psi^{1}_{R}\gamma^{\mu}\psi^{1}_{R}
-\bar{\psi}^{2}_{R}\gamma_{\mu}\psi^{2}_{R}\gamma^{\mu}\psi^{1}_{R}\right)=0\\
&i\gamma^{\mu}\nabla_{\mu}\psi^{2}_{L}
+\frac{3}{16}\left(\bar{\psi}^{1}_{L}\gamma_{\mu}\psi^{1}_{L}\gamma^{\mu}\psi^{2}_{L}
+\bar{\psi}^{2}_{L}\gamma_{\mu}\psi^{2}_{L}\gamma^{\mu}\psi^{2}_{L}
-\bar{\psi}^{1}_{R}\gamma_{\mu}\psi^{1}_{R}\gamma^{\mu}\psi^{2}_{L}
-\bar{\psi}^{2}_{R}\gamma_{\mu}\psi^{2}_{R}\gamma^{\mu}\psi^{2}_{L}\right)=0\\
&i\gamma^{\mu}\nabla_{\mu}\psi^{2}_{R}
-\frac{3}{16}\left(\bar{\psi}^{1}_{L}\gamma_{\mu}\psi^{1}_{L}\gamma^{\mu}\psi^{2}_{R}
+\bar{\psi}^{2}_{L}\gamma_{\mu}\psi^{2}_{L}\gamma^{\mu}\psi^{2}_{R}
-\bar{\psi}^{1}_{R}\gamma_{\mu}\psi^{1}_{R}\gamma^{\mu}\psi^{2}_{R}
-\bar{\psi}^{2}_{R}\gamma_{\mu}\psi^{2}_{R}\gamma^{\mu}\psi^{2}_{R}\right)=0
\end{eqnarray}
which may be written in the form
\begin{eqnarray}
&i\gamma^{\mu}\nabla_{\mu} L
-\frac{1}{2}g\vec{\sigma}\cdot\vec{A}_{\mu}\gamma^{\mu}L
-\frac{1}{6}g'B_{\mu}\gamma^{\mu}L-G_{u}i\sigma^{2}\phi^{\ast}u-G_{d}\phi d+O_{3}=0
\label{Leftquark}\\
&i\gamma^{\mu}\nabla_{\mu} u-\frac{2}{3}g'B_{\mu}\gamma^{\mu}u+G_{u}i\phi^{T}\sigma^{2}L+O_{3}=0
\label{rightuquark}\\
&i\gamma^{\mu}\nabla_{\mu} d+\frac{1}{3}g'B_{\mu}\gamma^{\mu}d-G_{d}\phi^{\dagger}L+O_{3}=0
\label{rightdquark}
\end{eqnarray}
when we define
\begin{eqnarray}
&\left(\psi^{1}_{R}\right)=u\ \ \ \ \left(\psi^{2}_{R}\right)=d\ \ \ \ \ \ \ \ 
\left(\begin{tabular}{c}$\psi^{1}_{L}$\\ $\psi^{2}_{L}$\end{tabular}\right)=L
\end{eqnarray}
with
\begin{eqnarray}
&\frac{15}{32G_{d}}\left(\bar{d}L\right)
+\frac{3i}{16G_{u}}\left(\bar{u}\sigma^{2}L\right)^{\ast}=\phi
\end{eqnarray}
and
\begin{eqnarray}
&\frac{9}{32}\left(\bar{L}\gamma_{\mu}\frac{\mathbb{I}}{2}L
+2\bar{u}\gamma_{\mu}u-\bar{d}\gamma_{\mu}d\right)+q\Gamma_{\mu}=g'B_{\mu}\\
&\frac{9}{32}\bar{L}\gamma_{\mu}\frac{\vec{\sigma}}{2}L=g\vec{A}_{\mu}
\end{eqnarray}
as new fields. In the field equations (\ref{Leftquark}), (\ref{rightuquark}) and (\ref{rightdquark}), the terms $O_{3}$ represent residual interactions written in the form of three-field vertices. Under the reasonable assumption for which these terms may be negligible, this is formally identical to the system of field equations for the quark fields before the symmetry breaking in the Standard Model.

The Abelian field $\Gamma_{\mu}$ must be fundamental and its coupling is given in terms of the $-\frac{1}{6}q$, $-\frac{2}{3}q$ and $\frac{1}{3}q$ charges for the left-handed $L$, right-handed $u$ and $d$ quarks respectively; the Abelian field, given by the current $\frac{9}{32}(\bar{L}\gamma_{\mu}\frac{\mathbb{I}}{2}L +2\bar{u}\gamma_{\mu}u -\bar{d}\gamma_{\mu}d)$, is composite in terms of those quarks and what allows the present prescription to work is that this current is defined once but its appearance occurs in terms of the $-\frac{1}{6}$, $-\frac{2}{3}$ and $\frac{1}{3}$ factors in front of the $L$, $u$ and $d$ quarks respectively: this is the reason for which the current $\frac{9}{32} (\bar{L}\gamma_{\mu}\frac{\mathbb{I}}{2}L+2\bar{u}\gamma_{\mu}u-\bar{d}\gamma_{\mu}d)$ and the Abelian field $q\Gamma_{\mu}$ may always be reabsorbed into the definition of the Abelian field $g'B_{\mu}$ still fundamental and still occurring with the $-\frac{1}{6}$, $-\frac{2}{3}$ and $\frac{1}{3}$ factors in front of the $L$, $u$ and $d$ quarks again respectively. By following the same method presented above, it is possible to establish the form of the transformation laws for the quarks, the scalar and vector fields before the symmetry breaking in the Standard Model.
\subsection{The massive spinor-semispinor system} 
In this final example, we shall deal with the case in which one of the two spinors is massive while the other is massless and taken to be a semispinor having only the left-handed projection \cite{F/1}. The aim is to compare this case with that of the electroweak interactions for leptons in a situation in which there is no symmetry breaking at all.

In the case is which one is a massive spinor and the other a massless semispinor having only the left-handed projection, the two different masses prevent them to mix forming doublets; this situation is such that the two fields are always asymmetric and they have to be identified with the electron and neutrino field right from the beginning. 

In this case the field equations for the electron and neutrino fields $e$ and $\nu$ are 
\begin{eqnarray}
&i\gamma^{\mu}\nabla_{\mu} e
-\frac{3}{16}\left(\overline{e}\gamma_{\mu}e\gamma^{\mu}e
+\overline{\nu}\gamma_{\mu}\nu\gamma^{\mu}\gamma_{5}e\right)-me=0\\
&i\gamma^{\mu}\nabla_{\mu}\nu
-\frac{3}{16}\overline{e}\gamma_{\mu}\gamma_{5}e\gamma^{\mu}\nu=0
\end{eqnarray}
which can be transformed into the following
\begin{eqnarray}
\label{equations}
&i\gamma^{\mu}\nabla_{\mu} e-\frac{3}{8}(\cos{\theta})^{2}\overline{e}\gamma^{\mu}e\gamma_{\mu}e
+q\tan{\theta}Z_{\mu}\gamma^{\mu}e
-\frac{g}{2\cos{\theta}}Z_{\mu}\gamma^{\mu}e_{L}+\frac{g}{\sqrt{2}}W^{*}_{\mu}\gamma^{\mu}\nu
-me=0\label{electron}\\
&i\gamma^{\mu}\nabla_{\mu}\nu+\frac{g}{2\cos{\theta}}Z_{\mu}\gamma^{\mu}\nu
+\frac{g}{\sqrt{2}}W_{\mu}\gamma^{\mu}e_{L}=0
\label{neutrino}
\end{eqnarray}
once we define
\begin{eqnarray}
&Z^{\mu}=-\left[2(\sin{\theta})^{2}\overline{e}\gamma^{\mu}e-\overline{e}_{L}\gamma^{\mu}e_{L}
+\overline{\nu}\gamma^{\mu}\nu\right]\left(\frac{3\cot{\theta}}{16q}\right)
\label{neutral}\\
&W^{\mu}=-\left(\overline{e}_{L}\gamma^{\mu}\nu\right)
\left[\frac{12(\sin{\theta})^{2}-3}{16q\sqrt{2}\sin{\theta}}\right]
\label{charged}
\end{eqnarray}
as the new fields.

The field equations (\ref{electron}) and (\ref{neutrino}) are formally identical to the system of field equations for the lepton fields after the symmetry breaking and mass generation in the Standard Model; however, there has been no generation of any mass nor breaking of any symmetry because there has never been any symmetry in the model while there has always been mass for the fermion fields. 

The lack of symmetry of the mediators comes from the fact that they are composite states of massive spinors due to the presence of torsion, undergoing to no transformation law. The lack of the standard Higgs field comes from the fact that, in this approach, it would be useless. These two features constitute important discrepancies with respect to the Standard Model: in fact, in this example, we must expect the weak bosons to display internal structure when the energy is high enough to probe their compositeness while in the Standard Model they must remain structureless at any energy scale; and because the present model predicts the absence of the Higgs field while the Standard Model needs its presence, then the detection of the Higgs or its persistent elusive character would settle the controversy definitively.

As an additional comment, we have to specify that, in this construction, there is a fundamental difference between the case of leptons and that of quarks, that is in the former, the weak interaction are reproduced exactly, whereas, in the latter, they have been reproduced only partially becoming exact when the three-field interactions are neglected: if, on the one hand, the corrections we have in the case of quarks may well be negligible, especially in situations in which also chromodynamics is present and where all quarks are confined as partons within nucleons, on the other hand there must be no correction for leptons since their interactions are relatively simple. In other words, we may say that in a comparison between the present approach and the Standard Model, we may have small deviations in situations in which the dynamics is dirty enough to keep them hidden, such as for quarks, but no deviation in situation the dynamics is clean enough to display them clearly, such as for leptons. Thus said, it is obvious how for our approach to work, the neutrino must be left-handed solely, and this means in particular massless. This situation may appear to be in contrast with what is becoming the accepted paradigm for which neutrinos appear instead to be massive; however the mass of neutrinos is seen as a consequence of the theory that links mass to oscillation and the observation of the oscillations themselves, the last being an irrefutable experimental evidence while the first being a theory that still has with many conceptual problems: these two problems may be solved at once by a model in which neutrino are massless and nevertheless still capable of oscillations. Incidentally, we do not have to look much far to find such a model, because in presence of torsion, neutrinos do oscillate precisely because they are massless, as discussed in \cite{F/2}.

Finally, we would like to spend a few words about the differences between the case in which fermions are initially massless yielding the dynamics of the Standard Model before the symmetry breaking, and that in which they are already massive yielding the dynamics of the Standard Model after the symmetry breaking; what we would like to specify is that, in this last instance, the field equations are those that correspond to that of the Standard Model after the symmetry breaking, not because we have witnessed the breakdown of an initial symmetry but rather because the system has never been symmetric. In this last example the problem of having a symmetry breaking will be circumvented.

These results show that torsion induces interactions among spinors whose form recalls that of the weak forces between fermions: what this implies is that for the Dirac fermionic field equations, the presence of torsion would be manifested disguised by weak forces: therefore the information about the weak forces is stored within torsion, in a similar way as the information about gravitation is stored within the metric; this suggests that the weak forces and gravity (although intrinsically different) may be accommodated into (different) parts of the same connection. This is not what we usually call unification, and yet it clearly is a stance of unification, inasmuch as a conceptual perspective is adopted.
\section{Conclusions}
We have shown that, starting from a theory of gravity formulated in 5-dimensions, the 4D-spacetime reduction procedure gives rise to extensions of GR in which the additional gravitational degrees of freedom can be considered as additional physical scalar fields. Such scalar fields are geometrical deformations undergoing the $GL(4)$-group of diffeomorphisms and are capable of generating the masses of particles. Furthermore, the $f(R)$ non-linearities encode energy-dependent running couplings able to fine-tune the scale. Finally, torsion gives rise to forces having the structure of weak force.
In other words, for Dirac fermionic field equations, the presence of torsion would be manifested in the guise of weak forces: this means that, contrary to what is commonly believed, torsion might not be a quantity we never observed due to its weakness, instead, it might actually be a quantity whose effects are constantly observed in weak experiments, although we were not ready to think weak forces as a long-range manifestation of torsion. If this were true, it would give rise to some sort of unification of weak and gravitational forces, as they may be included respectively in torsion and metric degrees of freedom, that is in different parts of the same connection. This connection is naturally accompanied by electrodynamics, within the most general spinorial connection of the spinor field. The only thing that would be left out of this approach would be chromodynamics. However, this does not constitute a bad point of the present scheme compared to the commonly accepted Standard Model: in fact, neither here nor in the Standard Model chromodynamics is unified with all other interactions, and actually in the Standard Model, the weak and electrodynamic interactions are not unified as well, since the $U(1)\times SU(2)$ group, and for that matter the full $U(1)\times SU(2)\times SU(3)$ group are \emph{no unification} of sub-groups into a larger group, but only a mere albeit quite elegant \emph{patching} of sub-groups. 

The philosophical implications of this approach are profound, as this would force us to rethink the issue of unification in physics under a conceptual perspective, rather than according to the usual mathematical symmetry-based way, a way that, despite the \emph{plethora} of possible symmetry groups, we have been investigating in the last forty years, has given no real advance since the $SU(5)$-model predicted the decay of the proton, yet to be discovered. In this prescription there is no need to postulate any new ingredient, because torsion is naturally part of the most general connection, the non-linearity of the $f(R)$ function comes out from the requirement that the action of gravitational field be not restricted to be the simplest Hilbert-Einstein action. If the action has to be written in terms of the Ricci scalar curvature in the most general way, then $f(R)$ where $R$ contains torsion does constitute the most general dynamics we may have for the gravitational background.

However, some crucial points have to be considered. On the one hand, torsion manufactures weak interactions, which, for fermions, are structurally identical to those arising from the Standard Model, and therefore any effect in the dynamics of fermions would be reproduced as in the Standard Model itself. Nevertheless, the gauge fields emerge as composite fields and, although at energies that are low enough not to probe their internal compositeness, they should look like those of the Standard Model. However, for higher energies discrepancies, the Standard Model must pop out. It is worth stressing that, in our approach, there is no standard Higgs field but a sort of gravitational Higgs mechanism is present. The validity of the presented scheme could be reasonably checked at LHC in short time, due to the increasing luminosities of the set-up. At the present stage, experiments are indicating, very preliminary, the presence of resonances and condensate states that cannot be considered final evidences for Higgs boson (at Fermi Lab, by CDF collaboration, and at CERN, in LHC, by CMS and ATLAS \cite{atlas}).

Incoming years will offer new data carrying information about the structure of the elementary constituents of our universe, and although the full confirmation of the Standard Model and its most accredited extensions are expected, nevertheless they may fail to come; if so, new aspects of the problem of unification in physics will have to be faced, and they may turn out to be even more exciting. And despite the direction taken since the $1960$s, the dream of unification in physics might be much more similar to what Einstein himself hoped it to be.

\end{document}